\documentclass[aps,longbibliography,twocolumn,showpacs,superscriptaddress,floatfix,prl]{revtex4-2}
\usepackage{amsfonts,hyperref}

\usepackage[T1]{fontenc}
\usepackage[utf8]{inputenc}

\usepackage{soul}
\usepackage{graphicx}
\usepackage{amsmath}
\usepackage{amssymb}
\usepackage{color}

\usepackage{bm}

\begin{document}
\title{Accelerating two-dimensional tensor network contractions using QR-decompositions}

\author{Yining \surname{Zhang}}
\thanks{Y.Z. and Q.Y. contributed equally to this work.}
\affiliation{Institute for Theoretical Physics, University of Amsterdam, Science Park 904, 1098 XH Amsterdam, The Netherlands}

 \author{Qi \surname{Yang}} 
 \thanks{Y.Z. and Q.Y. contributed equally to this work.}
 \affiliation{Institute for Theoretical Physics, University of Amsterdam, Science Park 904, 1098 XH Amsterdam, The Netherlands} 
 
 \author{Philippe \surname{Corboz}} \affiliation{Institute for Theoretical Physics, University of Amsterdam, Science Park 904, 1098 XH Amsterdam, The Netherlands}

\begin{abstract}
Infinite projected entangled-pair states (iPEPS) provide a powerful tool for studying strongly correlated systems directly in the thermodynamic limit. A core component of the algorithm is the approximate contraction of the iPEPS, where the computational bottleneck typically lies in the  singular value or eigenvalue decompositions involved in the renormalization step. This is particularly true on GPUs, where tensor contractions are substantially faster than these decompositions. Here we propose a contraction scheme for $C_{4v}$-symmetric tensor networks  based on combining the corner transfer matrix renormalization group (CTMRG) with QR-decompositions which are substantially faster -- especially on GPUs. Our approach achieves up to two orders of magnitude speedup  compared to  standard CTMRG without loss of accuracy and yields state-of-the-art results for the Heisenberg and $J_1$-$J_2$ models in less than one hour on an H100 GPU.

\end{abstract}
\maketitle

Tensor network methods have emerged as powerful tools for the study of quantum many-body systems, providing highly efficient   variational ans\"atze, with accuracy that can be systematically controlled by the bond dimension $D$. The most prominent example is the matrix product state (MPS), which has been extremely successful in capturing ground states of (quasi-)one-dimensional systems, forming the foundation of the density matrix renormalization group (DMRG) method~\cite{white1992,schollwoeck2011}. Its natural extension to two dimensions, known as projected entangled-pair states (PEPS)~\cite{verstraete2004,nishio2004,jordan2008}, allows for an efficient description of large two-dimensional systems, and even of systems directly in the thermodynamic limit with infinite PEPS (iPEPS).  These 2D tensor networks have found widespread application to challenging problems, ranging from frustrated spin systems to  strongly correlated electron systems, see, e.g., Refs.~\cite{corboz14_shastry, liao17, niesen17b, chen18, jahromi18, niesen18, yamaguchi18,kshetrimayum19b, chung19, lee20, gauthe20, jimenez21, czarnik21, hasik21, liu22b, hasik22, ponsioen23b, weerda24, hasik24,schmoll24,corboz25}.

A core part in 2D tensor network algorithms is the approximate contraction of the tensor network, whose accuracy is controlled by the bond dimension $\chi$. Various algorithms have been developed, including the corner transfer matrix renormalization group~\cite{nishino1996,nishino97, orus2009-1}, variational uniform MPS (VUMPS) algorithm~\cite{vanderstraeten15,zauner-stauber18,vanderstraeten22,zhang23}, (higher-order) tensor network renormalization group~\cite{levin07,zhao2010,xie12}, and tensor network renormalization~\cite{evenbly15}. Among those, CTMRG is the most used approach, including its generalizations~\cite{corboz2011,corboz14_tJ,fishman18}. Variants with reduced computational cost have been developed~\cite{xie17,haghshenas19,lan23,naumann25}, at the expense of introducing additional truncation steps controlled by an additional bond dimension $\chi'$ in the contraction scheme. The computational bottleneck of CTMRG lies typically in the involved singular value or eigenvalue decompositions. This is especially true on GPUs, where implementations of these decompositions are far less efficient than tensor contractions. Recently, the use of randomized SVD has been proposed for accelerated decompositions~\cite{richards25}, but its integration with energy minimization algorithms remains to be demonstrated.

In this Letter, we propose a variant of the CTMRG method for tensor networks that exhibit the $C_{4v}$
  symmetry of the underlying square lattice, %based on the QR decomposition. 
  based on a QR decomposition of a smaller part of the network than in standard CTMRG. 
  The main advantage is that the QR decomposition is significantly faster than the SVD or eigenvalue decomposition~\cite{unfried23} -- especially on GPUs -- paving the way for highly accelerated 2D tensor network calculations. We benchmark the approach for the 2D Heisenberg and $J_1$-$J_2$  models, where we observe a speed-up of one to two orders of magnitude compared to the standard CTMRG scheme. We reproduce and even outperform previous state-of-the-art results in about  one hour on an H100 GPU.

\begin{figure}[htb!]
  \centering
  \includegraphics[width=0.93\linewidth]{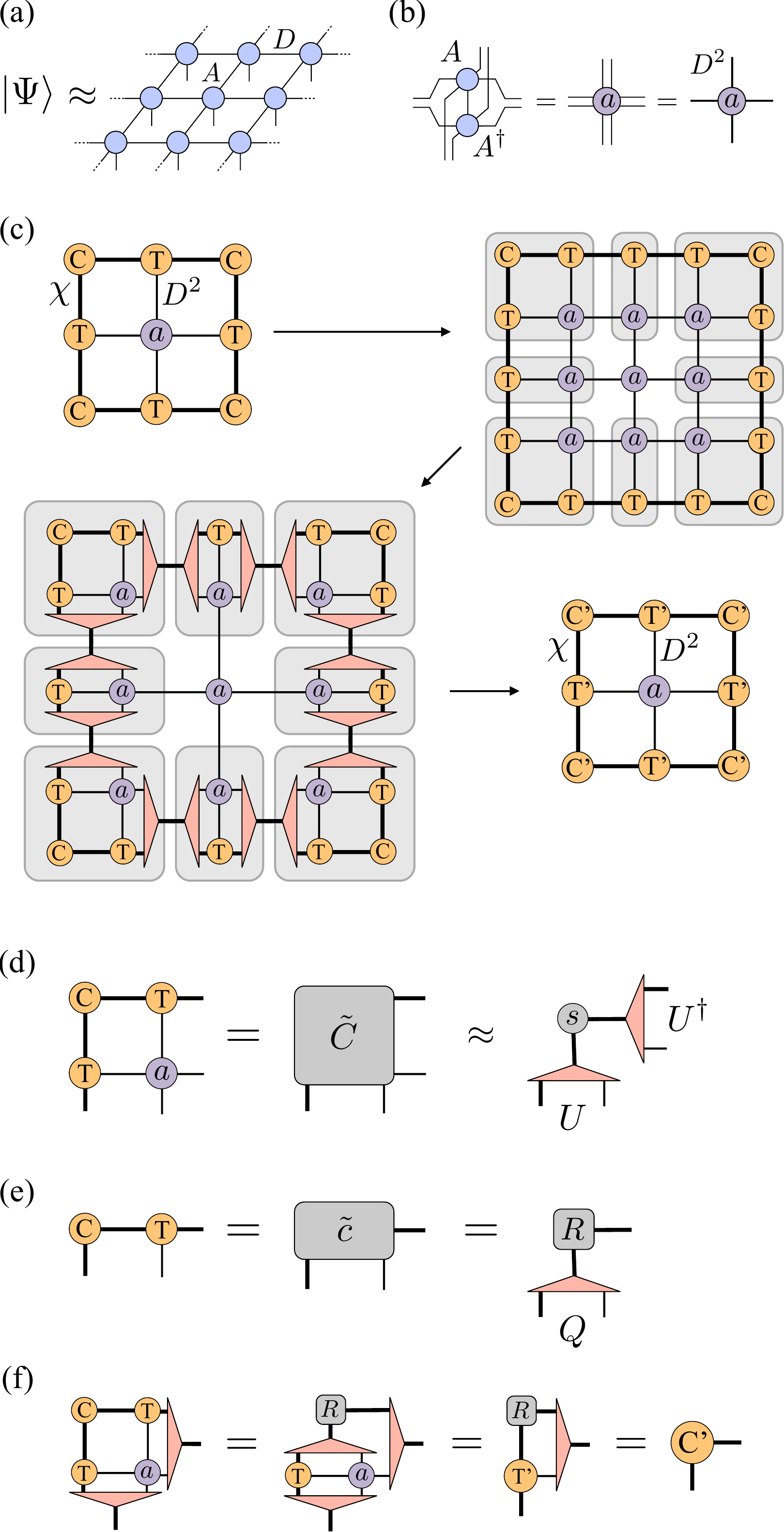}
  \caption{(a) Translationally invariant iPEPS represented by a network of rank-5 tensors $A$ with bond dimension $D$. (b)~Combining $A$ and $A^\dagger$ on each site results in tensor $a$ that can be reshaped into a rank-4 tensor with bond dimension $D^2$. The norm of a tensor network is represented by a square lattice network of the $a$ tensors. (c) CTMRG iteration in which the system is grown in all directions, followed by a renormalization step that truncates the bond dimension to $\chi$, resulting in updated boundary tensors $C'$, $T'$. (d) The renormalization step in CTMRG is based on an isometry $U$, obtained by diagonalizing an enlarged corner $\tilde{C}$ and retaining the $\chi$ eigenvectors with the largest magnitudes. (e) In QR-CTMRG, the renormalization is based on a smaller corner $\tilde{c}$, which is decomposed using a QR decomposition. The isometry $Q$ is then used in place of $U$ to perform the renormalization step. (f) Accelerated corner update based on the tensors $R$, $T'$, and $Q$, which is computationally cheaper than evaluating the standard diagram on the left-hand side.  }
  \label{fig1}
\end{figure}

\emph{iPEPS---}An iPEPS is a variational tensor network ansatz to represent quantum many-body states in 2D directly in the thermodynamic limit~\cite{verstraete2004, nishio2004, jordan2008}. Here we consider an ansatz made of a single rank-5, real-valued tensor $A$ repeated on a square lattice, as shown in Fig.~\ref{fig1}(a), where each tensor exhibits the  $C_{4v}$ symmetry of the lattice. Each tensor has one physical leg carrying the local Hilbert space of a lattice site of dimension $d$, and four auxiliary indices with bond dimension $D$ which connect to the neighboring tensors. Thus, each tensor has $d D^4$ elements, but the number of variational parameters is effectively reduced by exploiting the  $C_{4v}$ symmetry~\footnote{In the $C_{4v}$ symmetric case, the number of variational parameters is $n(D) = d/8 (D^4 + 2D^3 + 3D^2 + 2D)$ }. Contracting each tensor $A$ with its conjugate $A^\dagger$ (Fig.~\ref{fig1}(b))  results in a square lattice network of $a$ tensors, representing the norm of the state.

\emph{Standard CTMRG---}The CTMRG method is a powerful approach for systematically approximating the contraction of an infinite 2D tensor network~\cite{nishino1996}. The main idea is to start from an initial guess for the boundary tensors -- given by four corner tensors $C$ and four edge tensors $T$ -- and then let the system grow in all four directions by iteratively absorbing rows and columns of the network into the boundary tensors, see Fig.~\ref{fig1}(c)~\footnote{Here we initialize the boundary from the bulk tensors by contracting the corresponding bra- and ket- legs on the open boundary}. Each growth step increases the bond dimension $\chi$ of the boundary tensors by a factor of $D^2$, which is then truncated back to $\chi$ through a renormalization step. In standard CTMRG for the $C_{4v}$ symmetric case, the renormalization is performed using an isometry obtained by diagonalizing an extended corner $\tilde{C}$ (see Fig.~\ref{fig1}(d)), or alternatively by performing a singular value decomposition. The $\chi$ eigenvectors corresponding to the largest eigenvalues in magnitude (or singular values) are then retained. This choice is equivalent to the selection of relevant states used in DMRG~\cite{white1992}. The resulting isometry $U$ is then used to perform the renormalization step, yielding updated environment tensors $C'$ and $T'$. This procedure is repeated until convergence is reached. 
After convergence, the corner tensor $C$ and edge tensor $T$ effectively represent a quarter and a half-row (or half-column) of the infinite system, respectively, with an accuracy that is systematically controlled by $\chi$. 
The computationally most expensive part of CTMRG is the eigenvalue decomposition (or SVD), which scales as ${\cal O}(\chi^3 D^6)$ for a full decomposition. By using an iterative solver targeting only the $\chi$ dominant states, the cost is reduced to ${\cal O}(\chi^3 D^4)$ which matches the leading cost of the involved tensor contractions. However, due to a large prefactor, the decomposition step remains the computational bottleneck -- particularly on GPUs, where tensor contractions are extremely fast.

\emph{QR-CTMRG---}Here, we propose two key steps to substantially accelerate the computations. The first step is to use a lower-rank approximation $\tilde{c}$ of $\tilde{C}$, as shown in Fig.~\ref{fig1}(e), which effectively corresponds to reducing the system size in the vertical direction by one row~\footnote{Another type of reduced-rank approximation was introduced in Ref.~\cite{lan23} to decrease the computational cost. A reduced environment was also used in the directional CTMRG algorithm in Ref.~\cite{orus12} for the time evolution of a 1D quantum state, and in combination with a uniform MPS gauging method within a fixed-point CTMRG method~\cite{fishman18}}. That is, after $k$ steps, $\tilde{C}$ represents $k^2$ lattice sites, whereas $\tilde{c}$ represents only $k(k-1)$ lattice sites. An SVD of $\tilde{c}$ has a computational cost of only ${\cal O}(\chi^3 D^2)$.
Due to the reduced size of $\tilde c$, the renormalization step is expected to be less accurate in the first few iterations  than in standard CTMRG. However, in the limit of many iterations (i.e., large $k$), which is the relevant case for iPEPS, the error introduced by this approximation becomes negligible. In practice, we find that performing only a few (1-4) initial warm-up steps using the standard CTMRG scheme is sufficient. This warm-up phase also enables $\chi$ to be dynamically increased up to the desired target value. 
 
The second key ingredient is that, since $\tilde{c}$ already has a reduced rank of $\chi$, we can replace the SVD or eigenvalue decomposition with an (unpivoted) QR decomposition and use the resulting isometry $Q$ for the renormalization step. The advantage is that the QR decomposition is substantially faster, especially on GPUs, enabling highly accelerated iPEPS calculations.

Finally, we note that the renormalized corner tensor can be efficiently obtained from the contraction of  $T'$, $R$, and $Q$, as shown in the third diagram in Fig.~\ref{fig1}(f), with a computational cost that scales only as $\chi^3 D^2$. The leading scaling $\chi^3 D^4$  occurs only in the computation of $T'$, which is the dominant computational cost at large $D$ and $\chi$. A basic implementation of our algorithm is available on GitHub~\cite{GithubQRCTM}.

\emph{Automatic differentiation---}Determining the optimal variational parameters in iPEPS can be done either via imaginary time evolution~\cite{jordan2008,jiang2008,phien15} or by minimizing the variational energy~\cite{corboz16b,vanderstraeten16,liao19}. The most accurate approach currently relies on the latter, using automatic differentiation (AD) to compute the gradient of the energy with respect to the variational parameters~\cite{liao19}. An AD framework, offered by various libraries like PyTorch, JAX, Zygote, and others, automatically records the computation graph used to evaluate the energy. The gradient with respect to the input parameters is then obtained by backpropagating through the computation graph. The initial scheme required differentiating the full SVD, which increased the computational complexity   to ${\cal O}(\chi^3 D^6)$. More recently, a scheme based on iterative SVD or eigenvalue decomposition was introduced~\cite{francuz25}, resolving this problem. Since the QR decomposition is differentiable in an efficient way, the QR-CTMRG method can be naturally combined with AD-based optimization. As the optimizer, we use the L-BFGS algorithm, reusing the CTMRG boundary tensors from the previous step in each iteration, except during the line search of the algorithm. At each optimization step, we typically perform two CTMRG steps without gradient tracking, followed by $k$ steps with gradient tracking, where $k$ is chosen sufficiently large depending on $D$ and the model. As initial guess for the tensors at large $D$ we typically use an optimized tensor at lower $D$ combined with small random elements.

\emph{Results for the 2D Heisenberg model---}As a first test case, we consider the spin  $S=1/2$ Heisenberg model on a square lattice, defined by the Hamiltonian ${\cal H} = J \sum_{\langle i,j \rangle} \bm{S_i} \cdot \bm{S_j}$, exhibiting an antiferromagnetic ground state. By performing a local unitary transformation $e^{i\pi \sigma^y/2}$ on one of the sublattices, the ground state is mapped to a correlated ferromagnetic state that can be represented using a translationally invariant  and $C_{4v}$-symmetric iPEPS~\cite{hasik21,hasik24}.

\begin{figure}[tb]
  \centering
  \includegraphics[width=\linewidth]{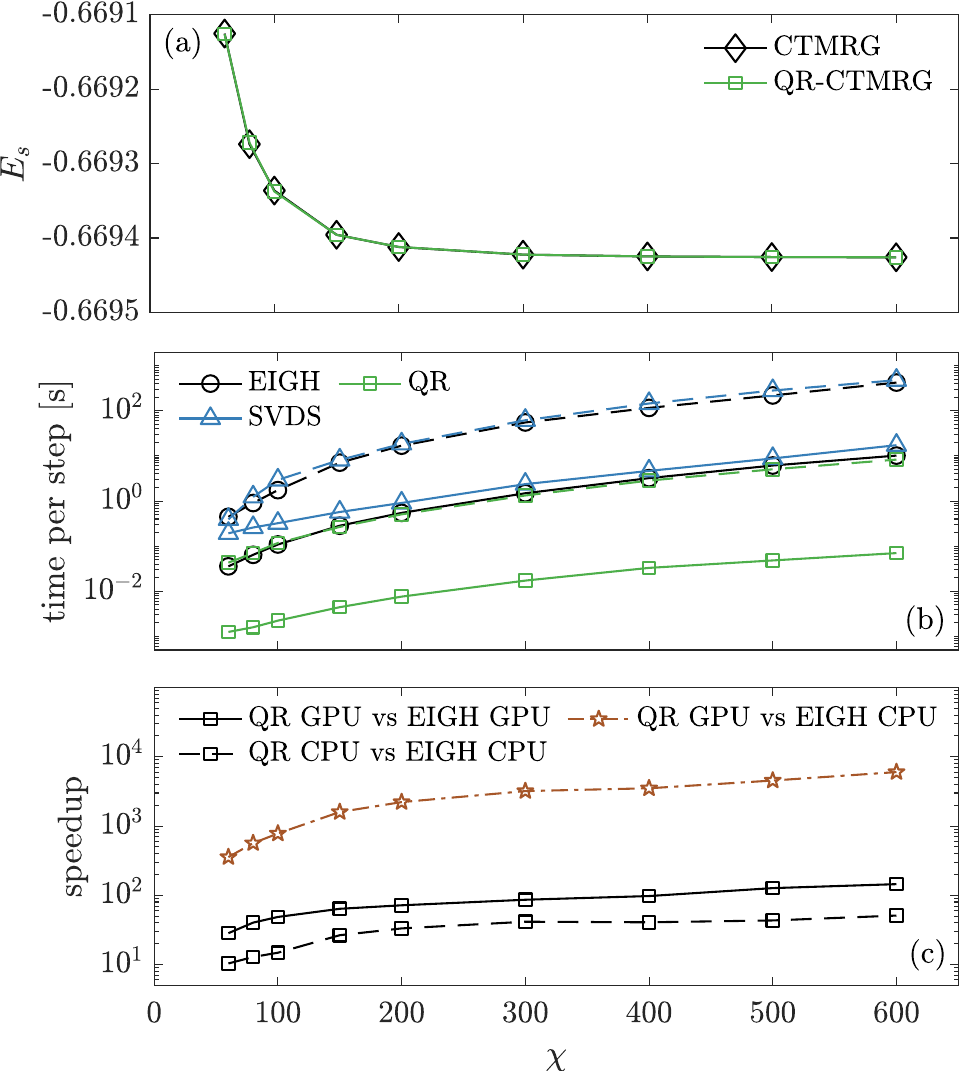}
  \caption{(a) Energy per site of the 2D Heisenberg model as a function of $\chi$ for $D=6$, obtained with standard and QR-CTMRG, showing similar convergence behavior.
(b) Average time per CTMRG step for the standard algorithm, using a Hermitian eigenvalue decomposition (EIGH) and an iterative SVD solver (SVDS), compared with QR-CTMRG (QR). Solid and dashed lines indicate results on an H100 GPU and on 8 cores on a CPU, respectively.
(c) Achieved speedup in computation time.
  }
  \label{fig:chiconv}
\end{figure}

We first focus on the contraction of an optimized $D=6$ iPEPS to compute the energy per site. Figure~\ref{fig:chiconv}(a) presents a comparison between standard CTMRG and QR-CTMRG, showing that both methods exhibit similar convergence behavior and accuracy as a function of~$\chi$. However, the latter is substantially faster, as demonstrated in Fig.~\ref{fig:chiconv}(b), which illustrates the average time per CTMRG step measured on 8 CPU cores and on an H100 GPU. 
At the largest value of $\chi = 600$, QR-CTMRG is approximately 50 times faster than standard CTMRG on the CPU, and 140 times faster on the GPU (see Fig.~\ref{fig:chiconv}(c)). The main reason for the speedup lies in the efficiency of the QR decomposition of the reduced corner~$\tilde c$: while in standard CTMRG 99\% of the time is spent on the eigenvalue decomposition, QR-CTMRG requires only a small fraction of time, typically less than 20\%, for the QR decomposition on the H100 GPU.
Comparing previous standard CTMRG calculations on CPUs with QR-CTMRG on GPUs reveals a speedup of more than three orders of magnitude, highlighting the enormous potential of QR-GPU-based 2D tensor network computations.

\begin{figure}[tb]
  \centering
  \includegraphics[width=\linewidth]{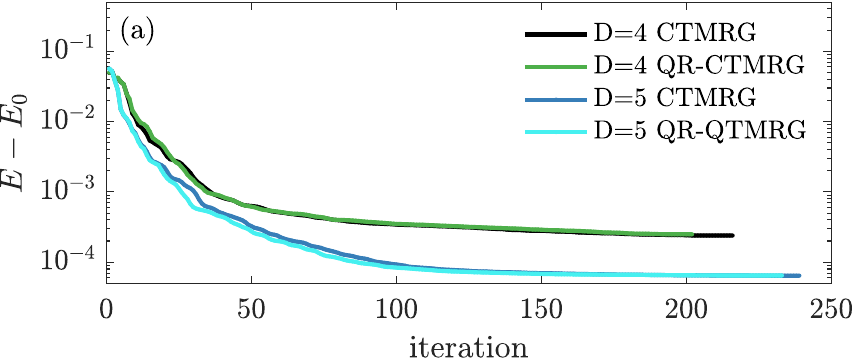} \\  \vspace{0.3cm}
    \includegraphics[width=0.96\linewidth]{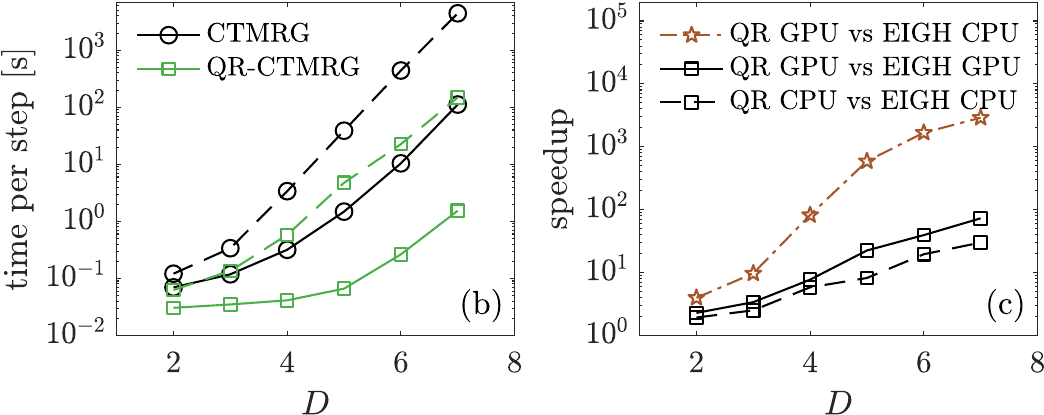}
  \caption{(a) Comparison of the optimization between standard and QR-CTMRG for the 2D Heisenberg model, showing that both methods converge in a similar fashion. Results are shown for $D=4$, $\chi=100$ and $D=5$, $\chi=150$. (b) Time per optimization step as a function of $D$ on a CPU (dashed lines) and GPU (solid lines), for $\chi = 5D^2$. One optimization step includes two CTMRG steps without gradient tracking, followed by $k$ CTMRG steps with gradient tracking, energy evaluation, and gradient computation. The number of steps $k$ is ${10,12,14,16,20,24}$ for $D = {2,\ldots,7}$. (c) Achieved speedup in computation time.
  }
  \label{fig:heis}
\end{figure}

We next test the method in the context of AD-based optimization. This represents a more challenging test case, as inaccuracies in the renormalization step can lead to errors in the gradient and negatively impact the optimization. However, we do not observe any disadvantages compared to standard CTMRG in this regard. Figure~\ref{fig:heis}(a) shows that the energy exhibits similar convergence behavior as a function of optimization steps, demonstrating the effectiveness of the approach when combined with AD. 
Figure~\ref{fig:heis}(b) presents a comparison of the time per optimization step as a function of $D$, and Fig.~\ref{fig:heis}(c) shows the corresponding speedup in computation time. Since the AD backward pass of the eigenvalue decomposition is faster than the forward pass, the overall speedup is smaller than in Fig.~\ref{fig:chiconv}(c). Nevertheless, a speedup by a factor greater than 70 can still be achieved at large $D$ on a GPU.

\begin{figure}[tb]
  \centering
  \includegraphics[width=\linewidth]{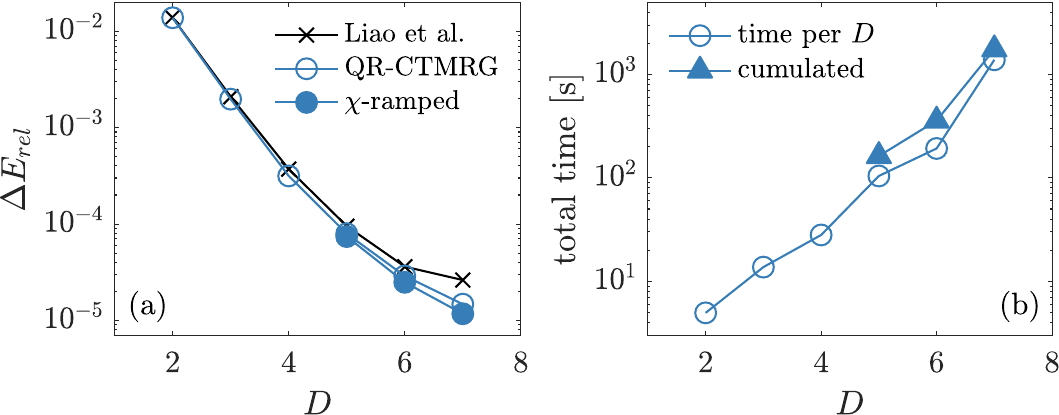} 
   \caption{(a) Relative error of the energy per site of the 2D Heisenberg model as a function of $D$, compared to previous data from Ref.~\cite{liao19}. The values of $\chi$ used for the optimization are $\chi_{opt}=\{30,80,100,200,200,300\}$ for $D=\{2,\ldots,7\}$. Solid symbols for $D=\{5,6,7\}$ show the energy evaluated at larger $\chi_{eval}=\{400,600,700\}$. (b) Total time of the simulations to obtain the data in (a) at each value of $D$. The solid symbols show the cumulated total time, including the final evaluation at large $\chi$, taking into account that the simulations were initialized from the converged state at the previous value of $D$ for $D>4$.
  }
  \label{fig:heisenergy}
\end{figure}

Finally, Fig.~\ref{fig:heisenergy}(a) presents the relative error of the optimized states using QR-CTMRG with respect to the extrapolated quantum Monte Carlo result from Ref.~\cite{Sandvik10}. For all values of $D$, we obtain an accuracy that is similar to or higher than previous iPEPS results from Refs.~\cite{liao19} and \cite{hasik24b}. The results are further refined by evaluating the final state at a larger boundary bond dimension ($\chi_{eval}$) than that used during the optimization ($\chi_{opt}$), as indicated in the figure caption.  Fig.~\ref{fig:heisenergy}(b) shows the corresponding total simulation time on an H100 GPU. The best result for $D = 7$ with an accuracy of 1.2e-5, which is currently the lowest variational energy in the thermodynamic limit, was obtained in only 30 minutes.

\begin{figure}[t]
  \centering
  \includegraphics[width=\linewidth]{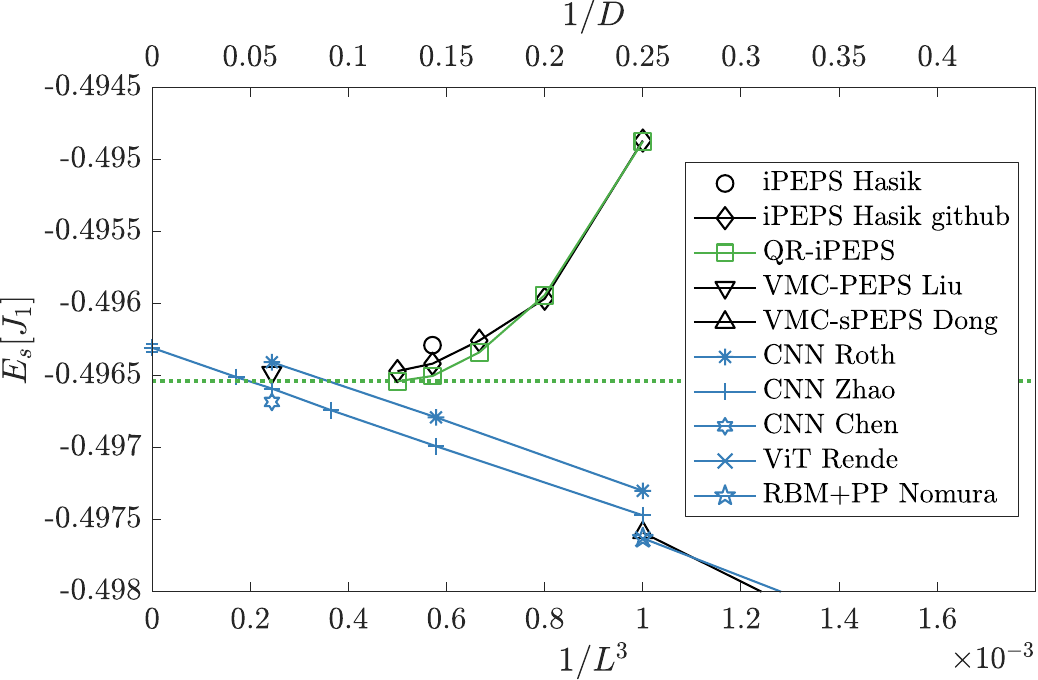} 
   \caption{QR based iPEPS results as a function of $1/D$ compared to previous iPEPS results and other variational results on finite system sizes. The dotted line is a guide to the eye showing our best result $E_s= -0.49654$ for $D=8$ with $\chi_{opt}=300$ and $\chi_{eval}=700$. }
  \label{fig:j1j2}
\end{figure}

\emph{Results for the $J_1$-$J_2$ Heisenberg model---}We now consider the substantially more challenging case of the $J_1$-$J_2$ Heisenberg model, which is out of reach for quantum Monte Carlo due to the negative sign problem. Several methods support the existence of an algebraic quantum spin liquid phase in the highly frustrated regime around $J_2/J_1=0.5$~\cite{hu13,richter15,wang18,hering19,hasik21,nomura21,liu22}. In Fig.~\ref{fig:j1j2} we present a comparison of the iPEPS variational energies with results from other variational methods, including previous iPEPS simulations~\cite{hasik21}, finite PEPS combined with VMC~\cite{dong25,liu24b}, and neural network quantum states, based on convolutional neural networks (CNN)~\cite{zhao22,liang23,roth23,chen24}, restricted Boltzmann machine combined with pair-product states (PP+RBM)~\cite{nomura21}, and deep vision transformer wave functions (ViT)~\cite{rende24}. 

Our QR-based iPEPS results for $D = 4, \ldots, 8$ are either comparable to or better than previous iPEPS results based on standard CTMRG~\cite{hasik21,hasik24b} (note that at large $D$ a lower $\chi$ was used  in the standard CTMRG approach). 
By pushing the $D = 8$ optimizations to $\chi = 300$ and evaluating at $\chi = 700$~\footnote{The boundary bond dimensions used here are   $\chi_{opt}=\{200, 200, 300, 300, 300\}$,  $\chi_{eval}=\{200, 200, 600, 600, 700\}$ for $D=\{4, 5, 6, 7, 8\}$, respectively.}, we achieve the currently lowest variational energy of $-0.49654$ in units of $J_1$ in the thermodynamic limit, within a total simulation time of 54 minutes on an H100 GPU. This value is considerably lower ($\sim 0.046\%$) than the extrapolated CNN result of $-0.49631(3)$~\cite{zhao22,liang23}, which was obtained using the whole Sunway supercomputer. A comparable energy can be reproduced with $D = 6, \chi_{opt}=300, \chi_{eval}=600$ in 12 minutes.

Several improved energies exist for finite system sizes without extrapolation to the thermodynamic limit. For example, the energy based on ViT~\cite{rende24} for a $10 \times 10$ system is $\sim 0.032\%$ lower than the CNN result, which represents a comparable improvement to QR-iPEPS in the thermodynamic limit, albeit with substantially higher computational time (4 days on 20 A100 GPUs) and a larger number of variational parameters (267'720 compared to 1'332 for $D = 8$).

\emph{Discussion---}We have demonstrated that QR-CTMRG offers a highly efficient approach for producing state-of-the-art iPEPS results in a remarkably short time on a GPU, even in the challenging case of the highly frustrated $J_1$-$J_2$ model.
The simulations can be pushed to even higher accuracy through longer computations, the use of multiple GPUs, and the exploitation of additional symmetries~\cite{singh2011,bauer2011,alphen25}. Further improvement may also be achieved through the fixed-point AD scheme~\cite{liao19}, which reduces memory usage and may enhance efficiency. We also note that QR-CTMRG  can also be applied to contract 2D tensor networks in the context of classical 2D partition functions~\cite{nishino1996}, or iPEPS at finite temperature using the variational tensor network renormalization (VTNR) approach~\cite{czarnik15b,czarnik16b,czarnik19}, enabling accelerated computation of thermodynamic quantities.

Our results demonstrate the great potential of QR-based contraction schemes on both CPUs and GPUs. While this work focused on the $C_{4v}$-symmetric case, a key open question is whether QR-based contraction schemes can also be developed for iPEPS with larger unit cells and without $C_{4v}$ symmetry, where the optimal projectors are no longer simple isometries~\cite{corboz14_tJ}. Exploring such schemes is an important direction for future research.

\acknowledgments
\emph{Acknowledgments -- } 
We thank Xiao Liang for providing the program execution time data. This project has received funding from the European Research Council (ERC) under the European Union's Horizon 2020 research and innovation programme (grant agreement No. 101001604). This work was carried out on the Dutch national e-infrastructure with the support of SURF Cooperative.

\bibliography{../../bib/refs.bib}

\begin{thebibliography}{79}%
\makeatletter
\providecommand \@ifxundefined [1]{%
 \@ifx{#1\undefined}
}%
\providecommand \@ifnum [1]{%
 \ifnum #1\expandafter \@firstoftwo
 \else \expandafter \@secondoftwo
 \fi
}%
\providecommand \@ifx [1]{%
 \ifx #1\expandafter \@firstoftwo
 \else \expandafter \@secondoftwo
 \fi
}%
\providecommand \natexlab [1]{#1}%
\providecommand \enquote  [1]{``#1''}%
\providecommand \bibnamefont  [1]{#1}%
\providecommand \bibfnamefont [1]{#1}%
\providecommand \citenamefont [1]{#1}%
\providecommand \href@noop [0]{\@secondoftwo}%
\providecommand \href [0]{\begingroup \@sanitize@url \@href}%
\providecommand \@href[1]{\@@startlink{#1}\@@href}%
\providecommand \@@href[1]{\endgroup#1\@@endlink}%
\providecommand \@sanitize@url [0]{\catcode `\\12\catcode `\$12\catcode
  `\&12\catcode `\#12\catcode `\^12\catcode `\_12\catcode `\%12\relax}%
\providecommand \@@startlink[1]{}%
\providecommand \@@endlink[0]{}%
\providecommand \url  [0]{\begingroup\@sanitize@url \@url }%
\providecommand \@url [1]{\endgroup\@href {#1}{\urlprefix }}%
\providecommand \urlprefix  [0]{URL }%
\providecommand \Eprint [0]{\href }%
\providecommand \doibase [0]{https://doi.org/}%
\providecommand \selectlanguage [0]{\@gobble}%
\providecommand \bibinfo  [0]{\@secondoftwo}%
\providecommand \bibfield  [0]{\@secondoftwo}%
\providecommand \translation [1]{[#1]}%
\providecommand \BibitemOpen [0]{}%
\providecommand \bibitemStop [0]{}%
\providecommand \bibitemNoStop [0]{.\EOS\space}%
\providecommand \EOS [0]{\spacefactor3000\relax}%
\providecommand \BibitemShut  [1]{\csname bibitem#1\endcsname}%
\let\auto@bib@innerbib\@empty
%</preamble>
\bibitem [{\citenamefont {White}(1992)}]{white1992}%
  \BibitemOpen
  \bibfield  {author} {\bibinfo {author} {\bibfnamefont {S.~R.}\ \bibnamefont
  {White}},\ }\bibfield  {title} {\bibinfo {title} {Density matrix formulation
  for quantum renormalization groups},\ }\href
  {https://doi.org/10.1103/PhysRevLett.69.2863} {\bibfield  {journal} {\bibinfo
   {journal} {Phys. Rev. Lett.}\ }\textbf {\bibinfo {volume} {69}},\ \bibinfo
  {pages} {2863} (\bibinfo {year} {1992})}\BibitemShut {NoStop}%
\bibitem [{\citenamefont {Schollw\"ock}(2011)}]{schollwoeck2011}%
  \BibitemOpen
  \bibfield  {author} {\bibinfo {author} {\bibfnamefont {U.}~\bibnamefont
  {Schollw\"ock}},\ }\bibfield  {title} {\bibinfo {title} {The density-matrix
  renormalization group in the age of matrix product states},\ }\href
  {https://doi.org/10.1016/j.aop.2010.09.012} {\bibfield  {journal} {\bibinfo
  {journal} {Annals of Physics}\ }\textbf {\bibinfo {volume} {326}},\ \bibinfo
  {pages} {96} (\bibinfo {year} {2011})}\BibitemShut {NoStop}%
\bibitem [{\citenamefont {Verstraete}\ and\ \citenamefont
  {Cirac}(2004)}]{verstraete2004}%
  \BibitemOpen
  \bibfield  {author} {\bibinfo {author} {\bibfnamefont {F.}~\bibnamefont
  {Verstraete}}\ and\ \bibinfo {author} {\bibfnamefont {J.~I.}\ \bibnamefont
  {Cirac}},\ }\bibfield  {title} {\bibinfo {title} {Renormalization algorithms
  for {Quantum}-{Many} {Body} {Systems} in two and higher dimensions},\ }\href
  {http://arxiv.org/abs/cond-mat/0407066} {\bibfield  {journal} {\bibinfo
  {journal} {arXiv:cond-mat/0407066}\ } (\bibinfo {year} {2004})}\BibitemShut
  {NoStop}%
\bibitem [{\citenamefont {Nishio}\ \emph {et~al.}(2004)\citenamefont {Nishio},
  \citenamefont {Maeshima}, \citenamefont {Gendiar},\ and\ \citenamefont
  {Nishino}}]{nishio2004}%
  \BibitemOpen
  \bibfield  {author} {\bibinfo {author} {\bibfnamefont {Y.}~\bibnamefont
  {Nishio}}, \bibinfo {author} {\bibfnamefont {N.}~\bibnamefont {Maeshima}},
  \bibinfo {author} {\bibfnamefont {A.}~\bibnamefont {Gendiar}},\ and\ \bibinfo
  {author} {\bibfnamefont {T.}~\bibnamefont {Nishino}},\ }\bibfield  {title}
  {\bibinfo {title} {{Tensor Product Variational Formulation for Quantum
  Systems}},\ }\href@noop {} {\bibfield  {journal} {\bibinfo  {journal}
  {Preprint}\ } (\bibinfo {year} {2004})},\ \Eprint
  {https://arxiv.org/abs/cond-mat/0401115} {arXiv:cond-mat/0401115}
  \BibitemShut {NoStop}%
\bibitem [{\citenamefont {Jordan}\ \emph {et~al.}(2008)\citenamefont {Jordan},
  \citenamefont {Or\'{u}s}, \citenamefont {Vidal}, \citenamefont {Verstraete},\
  and\ \citenamefont {Cirac}}]{jordan2008}%
  \BibitemOpen
  \bibfield  {author} {\bibinfo {author} {\bibfnamefont {J.}~\bibnamefont
  {Jordan}}, \bibinfo {author} {\bibfnamefont {R.}~\bibnamefont {Or\'{u}s}},
  \bibinfo {author} {\bibfnamefont {G.}~\bibnamefont {Vidal}}, \bibinfo
  {author} {\bibfnamefont {F.}~\bibnamefont {Verstraete}},\ and\ \bibinfo
  {author} {\bibfnamefont {J.~I.}\ \bibnamefont {Cirac}},\ }\bibfield  {title}
  {\bibinfo {title} {{Classical Simulation of Infinite-Size Quantum Lattice
  Systems in Two Spatial Dimensions}},\ }\href
  {https://doi.org/10.1103/PhysRevLett.101.250602} {\bibfield  {journal}
  {\bibinfo  {journal} {Phys. Rev. Lett.}\ }\textbf {\bibinfo {volume} {101}},\
  \bibinfo {pages} {250602} (\bibinfo {year} {2008})}\BibitemShut {NoStop}%
\bibitem [{\citenamefont {Corboz}\ and\ \citenamefont
  {Mila}(2014)}]{corboz14_shastry}%
  \BibitemOpen
  \bibfield  {author} {\bibinfo {author} {\bibfnamefont {P.}~\bibnamefont
  {Corboz}}\ and\ \bibinfo {author} {\bibfnamefont {F.}~\bibnamefont {Mila}},\
  }\bibfield  {title} {\bibinfo {title} {Crystals of bound states in the
  magnetization plateaus of the {S}hastry-{S}utherland model},\ }\href
  {https://doi.org/10.1103/PhysRevLett.112.147203} {\bibfield  {journal}
  {\bibinfo  {journal} {Phys. Rev. Lett.}\ }\textbf {\bibinfo {volume} {112}},\
  \bibinfo {pages} {147203} (\bibinfo {year} {2014})}\BibitemShut {NoStop}%
\bibitem [{\citenamefont {Liao}\ \emph {et~al.}(2017)\citenamefont {Liao},
  \citenamefont {Xie}, \citenamefont {Chen}, \citenamefont {Liu}, \citenamefont
  {Xie}, \citenamefont {Huang}, \citenamefont {Normand},\ and\ \citenamefont
  {Xiang}}]{liao17}%
  \BibitemOpen
  \bibfield  {author} {\bibinfo {author} {\bibfnamefont {H.~J.}\ \bibnamefont
  {Liao}}, \bibinfo {author} {\bibfnamefont {Z.~Y.}\ \bibnamefont {Xie}},
  \bibinfo {author} {\bibfnamefont {J.}~\bibnamefont {Chen}}, \bibinfo {author}
  {\bibfnamefont {Z.~Y.}\ \bibnamefont {Liu}}, \bibinfo {author} {\bibfnamefont
  {H.~D.}\ \bibnamefont {Xie}}, \bibinfo {author} {\bibfnamefont {R.~Z.}\
  \bibnamefont {Huang}}, \bibinfo {author} {\bibfnamefont {B.}~\bibnamefont
  {Normand}},\ and\ \bibinfo {author} {\bibfnamefont {T.}~\bibnamefont
  {Xiang}},\ }\bibfield  {title} {\bibinfo {title} {Gapless {Spin}-{Liquid}
  {Ground} {State} in the ${S}=1/2$ {Kagome} {Antiferromagnet}},\ }\href
  {https://doi.org/10.1103/PhysRevLett.118.137202} {\bibfield  {journal}
  {\bibinfo  {journal} {Phys. Rev. Lett.}\ }\textbf {\bibinfo {volume} {118}},\
  \bibinfo {pages} {137202} (\bibinfo {year} {2017})}\BibitemShut {NoStop}%
\bibitem [{\citenamefont {Niesen}\ and\ \citenamefont
  {Corboz}(2017)}]{niesen17b}%
  \BibitemOpen
  \bibfield  {author} {\bibinfo {author} {\bibfnamefont {I.}~\bibnamefont
  {Niesen}}\ and\ \bibinfo {author} {\bibfnamefont {P.}~\bibnamefont
  {Corboz}},\ }\bibfield  {title} {\bibinfo {title} {A tensor network study of
  the complete ground state phase diagram of the spin-1 bilinear-biquadratic
  {Heisenberg} model on the square lattice},\ }\href
  {https://doi.org/10.21468/SciPostPhys.3.4.030} {\bibfield  {journal}
  {\bibinfo  {journal} {SciPost Physics}\ }\textbf {\bibinfo {volume} {3}},\
  \bibinfo {pages} {030} (\bibinfo {year} {2017})}\BibitemShut {NoStop}%
\bibitem [{\citenamefont {Chen}\ \emph {et~al.}(2018)\citenamefont {Chen},
  \citenamefont {Vanderstraeten}, \citenamefont {Capponi},\ and\ \citenamefont
  {Poilblanc}}]{chen18}%
  \BibitemOpen
  \bibfield  {author} {\bibinfo {author} {\bibfnamefont {J.-Y.}\ \bibnamefont
  {Chen}}, \bibinfo {author} {\bibfnamefont {L.}~\bibnamefont
  {Vanderstraeten}}, \bibinfo {author} {\bibfnamefont {S.}~\bibnamefont
  {Capponi}},\ and\ \bibinfo {author} {\bibfnamefont {D.}~\bibnamefont
  {Poilblanc}},\ }\bibfield  {title} {\bibinfo {title} {Non-{Abelian} chiral
  spin liquid in a quantum antiferromagnet revealed by an {iPEPS} study},\
  }\href {https://doi.org/10.1103/PhysRevB.98.184409} {\bibfield  {journal}
  {\bibinfo  {journal} {Phys. Rev. B}\ }\textbf {\bibinfo {volume} {98}},\
  \bibinfo {pages} {184409} (\bibinfo {year} {2018})}\BibitemShut {NoStop}%
\bibitem [{\citenamefont {Jahromi}\ and\ \citenamefont
  {Or{\'u}s}(2018)}]{jahromi18}%
  \BibitemOpen
  \bibfield  {author} {\bibinfo {author} {\bibfnamefont {S.~S.}\ \bibnamefont
  {Jahromi}}\ and\ \bibinfo {author} {\bibfnamefont {R.}~\bibnamefont
  {Or{\'u}s}},\ }\bibfield  {title} {\bibinfo {title} {Spin-$\frac{1}{2}$
  {Heisenberg} antiferromagnet on the star lattice: {Competing}
  valence-bond-solid phases studied by means of tensor networks},\ }\href
  {https://doi.org/10.1103/PhysRevB.98.155108} {\bibfield  {journal} {\bibinfo
  {journal} {Phys. Rev. B}\ }\textbf {\bibinfo {volume} {98}},\ \bibinfo
  {pages} {155108} (\bibinfo {year} {2018})}\BibitemShut {NoStop}%
\bibitem [{\citenamefont {Niesen}\ and\ \citenamefont
  {Corboz}(2018)}]{niesen18}%
  \BibitemOpen
  \bibfield  {author} {\bibinfo {author} {\bibfnamefont {I.}~\bibnamefont
  {Niesen}}\ and\ \bibinfo {author} {\bibfnamefont {P.}~\bibnamefont
  {Corboz}},\ }\bibfield  {title} {\bibinfo {title} {Ground-state study of the
  spin-1 bilinear-biquadratic {Heisenberg} model on the triangular lattice
  using tensor networks},\ }\href {https://doi.org/10.1103/PhysRevB.97.245146}
  {\bibfield  {journal} {\bibinfo  {journal} {Phys. Rev. B}\ }\textbf {\bibinfo
  {volume} {97}},\ \bibinfo {pages} {245146} (\bibinfo {year}
  {2018})}\BibitemShut {NoStop}%
\bibitem [{\citenamefont {Yamaguchi}\ \emph {et~al.}(2018)\citenamefont
  {Yamaguchi}, \citenamefont {Sasaki}, \citenamefont {Okubo}, \citenamefont
  {Yoshida}, \citenamefont {Kida}, \citenamefont {Hagiwara}, \citenamefont
  {Kono}, \citenamefont {Kittaka}, \citenamefont {Sakakibara}, \citenamefont
  {Takigawa}, \citenamefont {Iwasaki},\ and\ \citenamefont
  {Hosokoshi}}]{yamaguchi18}%
  \BibitemOpen
  \bibfield  {author} {\bibinfo {author} {\bibfnamefont {H.}~\bibnamefont
  {Yamaguchi}}, \bibinfo {author} {\bibfnamefont {Y.}~\bibnamefont {Sasaki}},
  \bibinfo {author} {\bibfnamefont {T.}~\bibnamefont {Okubo}}, \bibinfo
  {author} {\bibfnamefont {M.}~\bibnamefont {Yoshida}}, \bibinfo {author}
  {\bibfnamefont {T.}~\bibnamefont {Kida}}, \bibinfo {author} {\bibfnamefont
  {M.}~\bibnamefont {Hagiwara}}, \bibinfo {author} {\bibfnamefont
  {Y.}~\bibnamefont {Kono}}, \bibinfo {author} {\bibfnamefont {S.}~\bibnamefont
  {Kittaka}}, \bibinfo {author} {\bibfnamefont {T.}~\bibnamefont {Sakakibara}},
  \bibinfo {author} {\bibfnamefont {M.}~\bibnamefont {Takigawa}}, \bibinfo
  {author} {\bibfnamefont {Y.}~\bibnamefont {Iwasaki}},\ and\ \bibinfo {author}
  {\bibfnamefont {Y.}~\bibnamefont {Hosokoshi}},\ }\bibfield  {title} {\bibinfo
  {title} {Field-enhanced quantum fluctuation in an ${S}=\frac{1}{2}$
  frustrated square lattice},\ }\href
  {https://doi.org/10.1103/PhysRevB.98.094402} {\bibfield  {journal} {\bibinfo
  {journal} {Phys. Rev. B}\ }\textbf {\bibinfo {volume} {98}},\ \bibinfo
  {pages} {094402} (\bibinfo {year} {2018})}\BibitemShut {NoStop}%
\bibitem [{\citenamefont {Kshetrimayum}\ \emph {et~al.}(2019)\citenamefont
  {Kshetrimayum}, \citenamefont {Balz}, \citenamefont {Lake},\ and\
  \citenamefont {Eisert}}]{kshetrimayum19b}%
  \BibitemOpen
  \bibfield  {author} {\bibinfo {author} {\bibfnamefont {A.}~\bibnamefont
  {Kshetrimayum}}, \bibinfo {author} {\bibfnamefont {C.}~\bibnamefont {Balz}},
  \bibinfo {author} {\bibfnamefont {B.}~\bibnamefont {Lake}},\ and\ \bibinfo
  {author} {\bibfnamefont {J.}~\bibnamefont {Eisert}},\ }\bibfield  {title}
  {\bibinfo {title} {Tensor network investigation of the double layer {Kagome}
  compound $\text{Ca}_{10}\text{Cr}_7\text{O}_{28}$},\ }\href
  {http://arxiv.org/abs/1904.00028} {\bibfield  {journal} {\bibinfo  {journal}
  {arXiv:1904.00028 [cond-mat, physics:quant-ph]}\ } (\bibinfo {year}
  {2019})}\BibitemShut {NoStop}%
\bibitem [{\citenamefont {Chung}\ and\ \citenamefont {Corboz}(2019)}]{chung19}%
  \BibitemOpen
  \bibfield  {author} {\bibinfo {author} {\bibfnamefont {S.~S.}\ \bibnamefont
  {Chung}}\ and\ \bibinfo {author} {\bibfnamefont {P.}~\bibnamefont {Corboz}},\
  }\bibfield  {title} {\bibinfo {title} {{SU(3)} fermions on the honeycomb
  lattice at $\frac{1}{3}$ filling},\ }\href
  {https://doi.org/10.1103/PhysRevB.100.035134} {\bibfield  {journal} {\bibinfo
   {journal} {Phys. Rev. B}\ }\textbf {\bibinfo {volume} {100}},\ \bibinfo
  {pages} {035134} (\bibinfo {year} {2019})}\BibitemShut {NoStop}%
\bibitem [{\citenamefont {Lee}\ \emph {et~al.}(2020)\citenamefont {Lee},
  \citenamefont {Kaneko}, \citenamefont {Chern}, \citenamefont {Okubo},
  \citenamefont {Yamaji}, \citenamefont {Kawashima},\ and\ \citenamefont
  {Kim}}]{lee20}%
  \BibitemOpen
  \bibfield  {author} {\bibinfo {author} {\bibfnamefont {H.-Y.}\ \bibnamefont
  {Lee}}, \bibinfo {author} {\bibfnamefont {R.}~\bibnamefont {Kaneko}},
  \bibinfo {author} {\bibfnamefont {L.~E.}\ \bibnamefont {Chern}}, \bibinfo
  {author} {\bibfnamefont {T.}~\bibnamefont {Okubo}}, \bibinfo {author}
  {\bibfnamefont {Y.}~\bibnamefont {Yamaji}}, \bibinfo {author} {\bibfnamefont
  {N.}~\bibnamefont {Kawashima}},\ and\ \bibinfo {author} {\bibfnamefont
  {Y.~B.}\ \bibnamefont {Kim}},\ }\bibfield  {title} {\bibinfo {title}
  {Magnetic field induced quantum phases in a tensor network study of {Kitaev}
  magnets},\ }\href {https://doi.org/10.1038/s41467-020-15320-x} {\bibfield
  {journal} {\bibinfo  {journal} {Nat. Comm.}\ }\textbf {\bibinfo {volume}
  {11}},\ \bibinfo {pages} {1639} (\bibinfo {year} {2020})}\BibitemShut
  {NoStop}%
\bibitem [{\citenamefont {Gauth{\'e}}\ \emph {et~al.}(2020)\citenamefont
  {Gauth{\'e}}, \citenamefont {Capponi}, \citenamefont {Mambrini},\ and\
  \citenamefont {Poilblanc}}]{gauthe20}%
  \BibitemOpen
  \bibfield  {author} {\bibinfo {author} {\bibfnamefont {O.}~\bibnamefont
  {Gauth{\'e}}}, \bibinfo {author} {\bibfnamefont {S.}~\bibnamefont {Capponi}},
  \bibinfo {author} {\bibfnamefont {M.}~\bibnamefont {Mambrini}},\ and\
  \bibinfo {author} {\bibfnamefont {D.}~\bibnamefont {Poilblanc}},\ }\bibfield
  {title} {\bibinfo {title} {Quantum spin liquid phases in the
  bilinear-biquadratic two-{SU}(4)-fermion {Hamiltonian} on the square
  lattice},\ }\href {https://doi.org/10.1103/PhysRevB.101.205144} {\bibfield
  {journal} {\bibinfo  {journal} {Phys. Rev. B}\ }\textbf {\bibinfo {volume}
  {101}},\ \bibinfo {pages} {205144} (\bibinfo {year} {2020})}\BibitemShut
  {NoStop}%
\bibitem [{\citenamefont {Jim{\'e}nez}\ \emph {et~al.}(2021)\citenamefont
  {Jim{\'e}nez}, \citenamefont {Crone}, \citenamefont {Fogh}, \citenamefont
  {Zayed}, \citenamefont {Lortz}, \citenamefont {Pomjakushina}, \citenamefont
  {Conder}, \citenamefont {L{\"a}uchli}, \citenamefont {Weber}, \citenamefont
  {Wessel}, \citenamefont {Honecker}, \citenamefont {Normand}, \citenamefont
  {R{\"u}egg}, \citenamefont {Corboz}, \citenamefont {R{\o}nnow},\ and\
  \citenamefont {Mila}}]{jimenez21}%
  \BibitemOpen
  \bibfield  {author} {\bibinfo {author} {\bibfnamefont {J.~L.}\ \bibnamefont
  {Jim{\'e}nez}}, \bibinfo {author} {\bibfnamefont {S.~P.~G.}\ \bibnamefont
  {Crone}}, \bibinfo {author} {\bibfnamefont {E.}~\bibnamefont {Fogh}},
  \bibinfo {author} {\bibfnamefont {M.~E.}\ \bibnamefont {Zayed}}, \bibinfo
  {author} {\bibfnamefont {R.}~\bibnamefont {Lortz}}, \bibinfo {author}
  {\bibfnamefont {E.}~\bibnamefont {Pomjakushina}}, \bibinfo {author}
  {\bibfnamefont {K.}~\bibnamefont {Conder}}, \bibinfo {author} {\bibfnamefont
  {A.~M.}\ \bibnamefont {L{\"a}uchli}}, \bibinfo {author} {\bibfnamefont
  {L.}~\bibnamefont {Weber}}, \bibinfo {author} {\bibfnamefont
  {S.}~\bibnamefont {Wessel}}, \bibinfo {author} {\bibfnamefont
  {A.}~\bibnamefont {Honecker}}, \bibinfo {author} {\bibfnamefont
  {B.}~\bibnamefont {Normand}}, \bibinfo {author} {\bibfnamefont
  {C.}~\bibnamefont {R{\"u}egg}}, \bibinfo {author} {\bibfnamefont
  {P.}~\bibnamefont {Corboz}}, \bibinfo {author} {\bibfnamefont {H.~M.}\
  \bibnamefont {R{\o}nnow}},\ and\ \bibinfo {author} {\bibfnamefont
  {F.}~\bibnamefont {Mila}},\ }\bibfield  {title} {\bibinfo {title} {A quantum
  magnetic analogue to the critical point of water},\ }\href
  {https://doi.org/10.1038/s41586-021-03411-8} {\bibfield  {journal} {\bibinfo
  {journal} {Nature}\ }\textbf {\bibinfo {volume} {592}},\ \bibinfo {pages}
  {370} (\bibinfo {year} {2021})}\BibitemShut {NoStop}%
\bibitem [{\citenamefont {Czarnik}\ \emph {et~al.}(2021)\citenamefont
  {Czarnik}, \citenamefont {Rams}, \citenamefont {Corboz},\ and\ \citenamefont
  {Dziarmaga}}]{czarnik21}%
  \BibitemOpen
  \bibfield  {author} {\bibinfo {author} {\bibfnamefont {P.}~\bibnamefont
  {Czarnik}}, \bibinfo {author} {\bibfnamefont {M.~M.}\ \bibnamefont {Rams}},
  \bibinfo {author} {\bibfnamefont {P.}~\bibnamefont {Corboz}},\ and\ \bibinfo
  {author} {\bibfnamefont {J.}~\bibnamefont {Dziarmaga}},\ }\bibfield  {title}
  {\bibinfo {title} {Tensor network study of the $m=\frac{1}{2}$ magnetization
  plateau in the {Shastry}-{Sutherland} model at finite temperature},\ }\href
  {https://doi.org/10.1103/PhysRevB.103.075113} {\bibfield  {journal} {\bibinfo
   {journal} {Phys. Rev. B}\ }\textbf {\bibinfo {volume} {103}},\ \bibinfo
  {pages} {075113} (\bibinfo {year} {2021})}\BibitemShut {NoStop}%
\bibitem [{\citenamefont {Hasik}\ \emph {et~al.}(2021)\citenamefont {Hasik},
  \citenamefont {Poilblanc},\ and\ \citenamefont {Becca}}]{hasik21}%
  \BibitemOpen
  \bibfield  {author} {\bibinfo {author} {\bibfnamefont {J.}~\bibnamefont
  {Hasik}}, \bibinfo {author} {\bibfnamefont {D.}~\bibnamefont {Poilblanc}},\
  and\ \bibinfo {author} {\bibfnamefont {F.}~\bibnamefont {Becca}},\ }\bibfield
   {title} {\bibinfo {title} {Investigation of the {N{\'e}el} phase of the
  frustrated {Heisenberg} antiferromagnet by differentiable symmetric tensor
  networks},\ }\href {https://doi.org/10.21468/SciPostPhys.10.1.012} {\bibfield
   {journal} {\bibinfo  {journal} {SciPost Physics}\ }\textbf {\bibinfo
  {volume} {10}},\ \bibinfo {pages} {012} (\bibinfo {year} {2021})}\BibitemShut
  {NoStop}%
\bibitem [{\citenamefont {Liu}\ \emph {et~al.}(2022{\natexlab{a}})\citenamefont
  {Liu}, \citenamefont {Hasik}, \citenamefont {Gong}, \citenamefont
  {Poilblanc}, \citenamefont {Chen},\ and\ \citenamefont {Gu}}]{liu22b}%
  \BibitemOpen
  \bibfield  {author} {\bibinfo {author} {\bibfnamefont {W.-Y.}\ \bibnamefont
  {Liu}}, \bibinfo {author} {\bibfnamefont {J.}~\bibnamefont {Hasik}}, \bibinfo
  {author} {\bibfnamefont {S.-S.}\ \bibnamefont {Gong}}, \bibinfo {author}
  {\bibfnamefont {D.}~\bibnamefont {Poilblanc}}, \bibinfo {author}
  {\bibfnamefont {W.-Q.}\ \bibnamefont {Chen}},\ and\ \bibinfo {author}
  {\bibfnamefont {Z.-C.}\ \bibnamefont {Gu}},\ }\bibfield  {title} {\bibinfo
  {title} {Emergence of {Gapless} {Quantum} {Spin} {Liquid} from {Deconfined}
  {Quantum} {Critical} {Point}},\ }\href
  {https://doi.org/10.1103/PhysRevX.12.031039} {\bibfield  {journal} {\bibinfo
  {journal} {Phys. Rev. X}\ }\textbf {\bibinfo {volume} {12}},\ \bibinfo
  {pages} {031039} (\bibinfo {year} {2022}{\natexlab{a}})}\BibitemShut
  {NoStop}%
\bibitem [{\citenamefont {Hasik}\ \emph {et~al.}(2022)\citenamefont {Hasik},
  \citenamefont {Van~Damme}, \citenamefont {Poilblanc},\ and\ \citenamefont
  {Vanderstraeten}}]{hasik22}%
  \BibitemOpen
  \bibfield  {author} {\bibinfo {author} {\bibfnamefont {J.}~\bibnamefont
  {Hasik}}, \bibinfo {author} {\bibfnamefont {M.}~\bibnamefont {Van~Damme}},
  \bibinfo {author} {\bibfnamefont {D.}~\bibnamefont {Poilblanc}},\ and\
  \bibinfo {author} {\bibfnamefont {L.}~\bibnamefont {Vanderstraeten}},\
  }\bibfield  {title} {\bibinfo {title} {Simulating {Chiral} {Spin} {Liquids}
  with {Projected} {Entangled}-{Pair} {States}},\ }\href
  {https://doi.org/10.1103/PhysRevLett.129.177201} {\bibfield  {journal}
  {\bibinfo  {journal} {Phys. Rev. Lett.}\ }\textbf {\bibinfo {volume} {129}},\
  \bibinfo {pages} {177201} (\bibinfo {year} {2022})}\BibitemShut {NoStop}%
\bibitem [{\citenamefont {Ponsioen}\ \emph {et~al.}(2023)\citenamefont
  {Ponsioen}, \citenamefont {Chung},\ and\ \citenamefont
  {Corboz}}]{ponsioen23b}%
  \BibitemOpen
  \bibfield  {author} {\bibinfo {author} {\bibfnamefont {B.}~\bibnamefont
  {Ponsioen}}, \bibinfo {author} {\bibfnamefont {S.~S.}\ \bibnamefont
  {Chung}},\ and\ \bibinfo {author} {\bibfnamefont {P.}~\bibnamefont
  {Corboz}},\ }\bibfield  {title} {\bibinfo {title} {Superconducting stripes in
  the hole-doped three-band {Hubbard} model},\ }\href
  {https://doi.org/10.1103/PhysRevB.108.205154} {\bibfield  {journal} {\bibinfo
   {journal} {Phys. Rev. B}\ }\textbf {\bibinfo {volume} {108}},\ \bibinfo
  {pages} {205154} (\bibinfo {year} {2023})}\BibitemShut {NoStop}%
\bibitem [{\citenamefont {Weerda}\ and\ \citenamefont
  {Rizzi}(2024)}]{weerda24}%
  \BibitemOpen
  \bibfield  {author} {\bibinfo {author} {\bibfnamefont {E.~L.}\ \bibnamefont
  {Weerda}}\ and\ \bibinfo {author} {\bibfnamefont {M.}~\bibnamefont {Rizzi}},\
  }\bibfield  {title} {\bibinfo {title} {Fractional quantum {Hall} states with
  variational projected entangled-pair states: {A} study of the bosonic
  {Harper}-{Hofstadter} model},\ }\href
  {https://doi.org/10.1103/PhysRevB.109.L241117} {\bibfield  {journal}
  {\bibinfo  {journal} {Phys. Rev. B}\ }\textbf {\bibinfo {volume} {109}},\
  \bibinfo {pages} {L241117} (\bibinfo {year} {2024})}\BibitemShut {NoStop}%
\bibitem [{\citenamefont {Hasik}\ and\ \citenamefont {Corboz}(2024)}]{hasik24}%
  \BibitemOpen
  \bibfield  {author} {\bibinfo {author} {\bibfnamefont {J.}~\bibnamefont
  {Hasik}}\ and\ \bibinfo {author} {\bibfnamefont {P.}~\bibnamefont {Corboz}},\
  }\bibfield  {title} {\bibinfo {title} {Incommensurate {Order} with
  {Translationally} {Invariant} {Projected} {Entangled}-{Pair} {States}:
  {Spiral} {States} and {Quantum} {Spin} {Liquid} on the {Anisotropic}
  {Triangular} {Lattice}},\ }\href
  {https://doi.org/10.1103/PhysRevLett.133.176502} {\bibfield  {journal}
  {\bibinfo  {journal} {Phys. Rev. Lett.}\ }\textbf {\bibinfo {volume} {133}},\
  \bibinfo {pages} {176502} (\bibinfo {year} {2024})}\BibitemShut {NoStop}%
\bibitem [{\citenamefont {Schmoll}\ \emph {et~al.}(2024)\citenamefont
  {Schmoll}, \citenamefont {Naumann}, \citenamefont {Weerda}, \citenamefont
  {Eisert},\ and\ \citenamefont {Iqbal}}]{schmoll24}%
  \BibitemOpen
  \bibfield  {author} {\bibinfo {author} {\bibfnamefont {P.}~\bibnamefont
  {Schmoll}}, \bibinfo {author} {\bibfnamefont {J.}~\bibnamefont {Naumann}},
  \bibinfo {author} {\bibfnamefont {E.~L.}\ \bibnamefont {Weerda}}, \bibinfo
  {author} {\bibfnamefont {J.}~\bibnamefont {Eisert}},\ and\ \bibinfo {author}
  {\bibfnamefont {Y.}~\bibnamefont {Iqbal}},\ }\bibfield  {title} {\bibinfo
  {title} {Bathing in a sea of candidate quantum spin liquids: {From} the
  gapless ruby to the gapped maple-leaf lattice},\ }\href
  {https://arxiv.org/abs/2407.07145v2} {\bibfield  {journal} {\bibinfo
  {journal} {arXiv:2407.07145 [cond-mat]}\ } (\bibinfo {year}
  {2024})}\BibitemShut {NoStop}%
\bibitem [{\citenamefont {Corboz}\ \emph {et~al.}(2025)\citenamefont {Corboz},
  \citenamefont {Zhang}, \citenamefont {Ponsioen},\ and\ \citenamefont
  {Mila}}]{corboz25}%
  \BibitemOpen
  \bibfield  {author} {\bibinfo {author} {\bibfnamefont {P.}~\bibnamefont
  {Corboz}}, \bibinfo {author} {\bibfnamefont {Y.}~\bibnamefont {Zhang}},
  \bibinfo {author} {\bibfnamefont {B.}~\bibnamefont {Ponsioen}},\ and\
  \bibinfo {author} {\bibfnamefont {F.}~\bibnamefont {Mila}},\ }\bibfield
  {title} {\bibinfo {title} {Quantum spin liquid phase in the
  {Shastry}-{Sutherland} model revealed by high-precision infinite projected
  entangled-pair states},\ }\bibfield  {journal} {\bibinfo  {journal}
  {arXiv:2502.14091 [cond-mat]}\ }\href
  {https://doi.org/10.48550/arXiv.2502.14091} {10.48550/arXiv.2502.14091}
  (\bibinfo {year} {2025})\BibitemShut {NoStop}%
\bibitem [{\citenamefont {Nishino}\ and\ \citenamefont
  {Okunishi}(1996)}]{nishino1996}%
  \BibitemOpen
  \bibfield  {author} {\bibinfo {author} {\bibfnamefont {T.}~\bibnamefont
  {Nishino}}\ and\ \bibinfo {author} {\bibfnamefont {K.}~\bibnamefont
  {Okunishi}},\ }\bibfield  {title} {\bibinfo {title} {{Corner Transfer Matrix
  Renormalization Group Method}},\ }\href {https://doi.org/10.1143/JPSJ.65.891}
  {\bibfield  {journal} {\bibinfo  {journal} {J. Phys. Soc. Jpn.}\ }\textbf
  {\bibinfo {volume} {65}},\ \bibinfo {pages} {891} (\bibinfo {year}
  {1996})}\BibitemShut {NoStop}%
\bibitem [{\citenamefont {Nishino}\ and\ \citenamefont
  {Okunishi}(1997)}]{nishino97}%
  \BibitemOpen
  \bibfield  {author} {\bibinfo {author} {\bibfnamefont {T.}~\bibnamefont
  {Nishino}}\ and\ \bibinfo {author} {\bibfnamefont {K.}~\bibnamefont
  {Okunishi}},\ }\bibfield  {title} {\bibinfo {title} {Corner {Transfer}
  {Matrix} {Algorithm} for {Classical} {Renormalization} {Group}},\ }\href
  {https://doi.org/10.1143/JPSJ.66.3040} {\bibfield  {journal} {\bibinfo
  {journal} {J. Phys. Soc. Jpn.}\ }\textbf {\bibinfo {volume} {66}},\ \bibinfo
  {pages} {3040} (\bibinfo {year} {1997})}\BibitemShut {NoStop}%
\bibitem [{\citenamefont {Or\'us}\ and\ \citenamefont
  {Vidal}(2009)}]{orus2009-1}%
  \BibitemOpen
  \bibfield  {author} {\bibinfo {author} {\bibfnamefont {R.}~\bibnamefont
  {Or\'us}}\ and\ \bibinfo {author} {\bibfnamefont {G.}~\bibnamefont {Vidal}},\
  }\bibfield  {title} {\bibinfo {title} {Simulation of two-dimensional quantum
  systems on an infinite lattice revisited: Corner transfer matrix for tensor
  contraction},\ }\href {https://doi.org/10.1103/PhysRevB.80.094403} {\bibfield
   {journal} {\bibinfo  {journal} {Phys. Rev. B}\ }\textbf {\bibinfo {volume}
  {80}},\ \bibinfo {pages} {094403} (\bibinfo {year} {2009})}\BibitemShut
  {NoStop}%
\bibitem [{\citenamefont {Vanderstraeten}\ \emph {et~al.}(2015)\citenamefont
  {Vanderstraeten}, \citenamefont {Mari\"en}, \citenamefont {Verstraete},\ and\
  \citenamefont {Haegeman}}]{vanderstraeten15}%
  \BibitemOpen
  \bibfield  {author} {\bibinfo {author} {\bibfnamefont {L.}~\bibnamefont
  {Vanderstraeten}}, \bibinfo {author} {\bibfnamefont {M.}~\bibnamefont
  {Mari\"en}}, \bibinfo {author} {\bibfnamefont {F.}~\bibnamefont
  {Verstraete}},\ and\ \bibinfo {author} {\bibfnamefont {J.}~\bibnamefont
  {Haegeman}},\ }\bibfield  {title} {\bibinfo {title} {Excitations and the
  tangent space of projected entangled-pair states},\ }\href
  {https://doi.org/10.1103/PhysRevB.92.201111} {\bibfield  {journal} {\bibinfo
  {journal} {Phys. Rev. B}\ }\textbf {\bibinfo {volume} {92}},\ \bibinfo
  {pages} {201111} (\bibinfo {year} {2015})}\BibitemShut {NoStop}%
\bibitem [{\citenamefont {Zauner-Stauber}\ \emph {et~al.}(2018)\citenamefont
  {Zauner-Stauber}, \citenamefont {Vanderstraeten}, \citenamefont {Fishman},
  \citenamefont {Verstraete},\ and\ \citenamefont
  {Haegeman}}]{zauner-stauber18}%
  \BibitemOpen
  \bibfield  {author} {\bibinfo {author} {\bibfnamefont {V.}~\bibnamefont
  {Zauner-Stauber}}, \bibinfo {author} {\bibfnamefont {L.}~\bibnamefont
  {Vanderstraeten}}, \bibinfo {author} {\bibfnamefont {M.~T.}\ \bibnamefont
  {Fishman}}, \bibinfo {author} {\bibfnamefont {F.}~\bibnamefont
  {Verstraete}},\ and\ \bibinfo {author} {\bibfnamefont {J.}~\bibnamefont
  {Haegeman}},\ }\bibfield  {title} {\bibinfo {title} {Variational optimization
  algorithms for uniform matrix product states},\ }\href
  {https://doi.org/10.1103/PhysRevB.97.045145} {\bibfield  {journal} {\bibinfo
  {journal} {Phys. Rev. B}\ }\textbf {\bibinfo {volume} {97}},\ \bibinfo
  {pages} {045145} (\bibinfo {year} {2018})}\BibitemShut {NoStop}%
\bibitem [{\citenamefont {Vanderstraeten}\ \emph {et~al.}(2022)\citenamefont
  {Vanderstraeten}, \citenamefont {Burgelman}, \citenamefont {Ponsioen},
  \citenamefont {Van~Damme}, \citenamefont {Vanhecke}, \citenamefont {Corboz},
  \citenamefont {Haegeman},\ and\ \citenamefont
  {Verstraete}}]{vanderstraeten22}%
  \BibitemOpen
  \bibfield  {author} {\bibinfo {author} {\bibfnamefont {L.}~\bibnamefont
  {Vanderstraeten}}, \bibinfo {author} {\bibfnamefont {L.}~\bibnamefont
  {Burgelman}}, \bibinfo {author} {\bibfnamefont {B.}~\bibnamefont {Ponsioen}},
  \bibinfo {author} {\bibfnamefont {M.}~\bibnamefont {Van~Damme}}, \bibinfo
  {author} {\bibfnamefont {B.}~\bibnamefont {Vanhecke}}, \bibinfo {author}
  {\bibfnamefont {P.}~\bibnamefont {Corboz}}, \bibinfo {author} {\bibfnamefont
  {J.}~\bibnamefont {Haegeman}},\ and\ \bibinfo {author} {\bibfnamefont
  {F.}~\bibnamefont {Verstraete}},\ }\bibfield  {title} {\bibinfo {title}
  {Variational methods for contracting projected entangled-pair states},\
  }\href {https://doi.org/10.1103/PhysRevB.105.195140} {\bibfield  {journal}
  {\bibinfo  {journal} {Phys. Rev. B}\ }\textbf {\bibinfo {volume} {105}},\
  \bibinfo {pages} {195140} (\bibinfo {year} {2022})}\BibitemShut {NoStop}%
\bibitem [{\citenamefont {Zhang}\ \emph {et~al.}()\citenamefont {Zhang},
  \citenamefont {Liang}, \citenamefont {Liao}, \citenamefont {Li},\ and\
  \citenamefont {Wang}}]{zhang23}%
  \BibitemOpen
  \bibfield  {author} {\bibinfo {author} {\bibfnamefont {X.-Y.}\ \bibnamefont
  {Zhang}}, \bibinfo {author} {\bibfnamefont {S.}~\bibnamefont {Liang}},
  \bibinfo {author} {\bibfnamefont {H.-J.}\ \bibnamefont {Liao}}, \bibinfo
  {author} {\bibfnamefont {W.}~\bibnamefont {Li}},\ and\ \bibinfo {author}
  {\bibfnamefont {L.}~\bibnamefont {Wang}},\ }\bibfield  {title} {\bibinfo
  {title} {Differentiable programming tensor networks for kitaev magnets},\
  }\href {https://doi.org/10.1103/PhysRevB.108.085103} {\bibfield  {journal}
  {\bibinfo  {journal} {Phys. Rev. B}\ }\textbf {\bibinfo {volume} {108}},\
  \bibinfo {pages} {085103}}\BibitemShut {NoStop}%
\bibitem [{\citenamefont {Levin}\ and\ \citenamefont {Nave}(2007)}]{levin07}%
  \BibitemOpen
  \bibfield  {author} {\bibinfo {author} {\bibfnamefont {M.}~\bibnamefont
  {Levin}}\ and\ \bibinfo {author} {\bibfnamefont {C.~P.}\ \bibnamefont
  {Nave}},\ }\bibfield  {title} {\bibinfo {title} {Tensor {Renormalization}
  {Group} {Approach} to {Two}-{Dimensional} {Classical} {Lattice} {Models}},\
  }\href {https://doi.org/10.1103/PhysRevLett.99.120601} {\bibfield  {journal}
  {\bibinfo  {journal} {Phys. Rev. Lett.}\ }\textbf {\bibinfo {volume} {99}},\
  \bibinfo {pages} {120601} (\bibinfo {year} {2007})}\BibitemShut {NoStop}%
\bibitem [{\citenamefont {Zhao}\ \emph {et~al.}(2010)\citenamefont {Zhao},
  \citenamefont {Xie}, \citenamefont {Chen}, \citenamefont {Wei}, \citenamefont
  {Cai},\ and\ \citenamefont {Xiang}}]{zhao2010}%
  \BibitemOpen
  \bibfield  {author} {\bibinfo {author} {\bibfnamefont {H.~H.}\ \bibnamefont
  {Zhao}}, \bibinfo {author} {\bibfnamefont {Z.~Y.}\ \bibnamefont {Xie}},
  \bibinfo {author} {\bibfnamefont {Q.~N.}\ \bibnamefont {Chen}}, \bibinfo
  {author} {\bibfnamefont {Z.~C.}\ \bibnamefont {Wei}}, \bibinfo {author}
  {\bibfnamefont {J.~W.}\ \bibnamefont {Cai}},\ and\ \bibinfo {author}
  {\bibfnamefont {T.}~\bibnamefont {Xiang}},\ }\bibfield  {title} {\bibinfo
  {title} {Renormalization of tensor-network states},\ }\href
  {https://doi.org/10.1103/PhysRevB.81.174411} {\bibfield  {journal} {\bibinfo
  {journal} {Phys. Rev. B}\ }\textbf {\bibinfo {volume} {81}},\ \bibinfo
  {pages} {174411} (\bibinfo {year} {2010})}\BibitemShut {NoStop}%
\bibitem [{\citenamefont {Xie}\ \emph {et~al.}(2012)\citenamefont {Xie},
  \citenamefont {Chen}, \citenamefont {Qin}, \citenamefont {Zhu}, \citenamefont
  {Yang},\ and\ \citenamefont {Xiang}}]{xie12}%
  \BibitemOpen
  \bibfield  {author} {\bibinfo {author} {\bibfnamefont {Z.~Y.}\ \bibnamefont
  {Xie}}, \bibinfo {author} {\bibfnamefont {J.}~\bibnamefont {Chen}}, \bibinfo
  {author} {\bibfnamefont {M.~P.}\ \bibnamefont {Qin}}, \bibinfo {author}
  {\bibfnamefont {J.~W.}\ \bibnamefont {Zhu}}, \bibinfo {author} {\bibfnamefont
  {L.~P.}\ \bibnamefont {Yang}},\ and\ \bibinfo {author} {\bibfnamefont
  {T.}~\bibnamefont {Xiang}},\ }\bibfield  {title} {\bibinfo {title}
  {Coarse-graining renormalization by higher-order singular value
  decomposition},\ }\href {https://doi.org/10.1103/PhysRevB.86.045139}
  {\bibfield  {journal} {\bibinfo  {journal} {Phys. Rev. B}\ }\textbf {\bibinfo
  {volume} {86}},\ \bibinfo {pages} {045139} (\bibinfo {year}
  {2012})}\BibitemShut {NoStop}%
\bibitem [{\citenamefont {Evenbly}\ and\ \citenamefont
  {Vidal}(2015)}]{evenbly15}%
  \BibitemOpen
  \bibfield  {author} {\bibinfo {author} {\bibfnamefont {G.}~\bibnamefont
  {Evenbly}}\ and\ \bibinfo {author} {\bibfnamefont {G.}~\bibnamefont
  {Vidal}},\ }\bibfield  {title} {\bibinfo {title} {Tensor {Network}
  {Renormalization}},\ }\href {https://doi.org/10.1103/PhysRevLett.115.180405}
  {\bibfield  {journal} {\bibinfo  {journal} {Phys. Rev. Lett.}\ }\textbf
  {\bibinfo {volume} {115}},\ \bibinfo {pages} {180405} (\bibinfo {year}
  {2015})}\BibitemShut {NoStop}%
\bibitem [{\citenamefont {Corboz}\ \emph {et~al.}(2011)\citenamefont {Corboz},
  \citenamefont {White}, \citenamefont {Vidal},\ and\ \citenamefont
  {Troyer}}]{corboz2011}%
  \BibitemOpen
  \bibfield  {author} {\bibinfo {author} {\bibfnamefont {P.}~\bibnamefont
  {Corboz}}, \bibinfo {author} {\bibfnamefont {S.~R.}\ \bibnamefont {White}},
  \bibinfo {author} {\bibfnamefont {G.}~\bibnamefont {Vidal}},\ and\ \bibinfo
  {author} {\bibfnamefont {M.}~\bibnamefont {Troyer}},\ }\bibfield  {title}
  {\bibinfo {title} {Stripes in the two-dimensional {t-J} model with infinite
  projected entangled-pair states},\ }\href
  {https://doi.org/10.1103/PhysRevB.84.041108} {\bibfield  {journal} {\bibinfo
  {journal} {Phys. Rev. B}\ }\textbf {\bibinfo {volume} {84}},\ \bibinfo
  {pages} {041108(R)} (\bibinfo {year} {2011})}\BibitemShut {NoStop}%
\bibitem [{\citenamefont {Corboz}\ \emph {et~al.}(2014)\citenamefont {Corboz},
  \citenamefont {Rice},\ and\ \citenamefont {Troyer}}]{corboz14_tJ}%
  \BibitemOpen
  \bibfield  {author} {\bibinfo {author} {\bibfnamefont {P.}~\bibnamefont
  {Corboz}}, \bibinfo {author} {\bibfnamefont {T.~M.}\ \bibnamefont {Rice}},\
  and\ \bibinfo {author} {\bibfnamefont {M.}~\bibnamefont {Troyer}},\
  }\bibfield  {title} {\bibinfo {title} {Competing {States} in the t-{J}
  {Model}: {Uniform} d-{Wave} {State} versus {Stripe} {State}},\ }\href
  {https://doi.org/10.1103/PhysRevLett.113.046402} {\bibfield  {journal}
  {\bibinfo  {journal} {Phys. Rev. Lett.}\ }\textbf {\bibinfo {volume} {113}},\
  \bibinfo {pages} {046402} (\bibinfo {year} {2014})}\BibitemShut {NoStop}%
\bibitem [{\citenamefont {Fishman}\ \emph {et~al.}(2018)\citenamefont
  {Fishman}, \citenamefont {Vanderstraeten}, \citenamefont {Zauner-Stauber},
  \citenamefont {Haegeman},\ and\ \citenamefont {Verstraete}}]{fishman18}%
  \BibitemOpen
  \bibfield  {author} {\bibinfo {author} {\bibfnamefont {M.~T.}\ \bibnamefont
  {Fishman}}, \bibinfo {author} {\bibfnamefont {L.}~\bibnamefont
  {Vanderstraeten}}, \bibinfo {author} {\bibfnamefont {V.}~\bibnamefont
  {Zauner-Stauber}}, \bibinfo {author} {\bibfnamefont {J.}~\bibnamefont
  {Haegeman}},\ and\ \bibinfo {author} {\bibfnamefont {F.}~\bibnamefont
  {Verstraete}},\ }\bibfield  {title} {\bibinfo {title} {Faster methods for
  contracting infinite two-dimensional tensor networks},\ }\href
  {https://doi.org/10.1103/PhysRevB.98.235148} {\bibfield  {journal} {\bibinfo
  {journal} {Phys. Rev. B}\ }\textbf {\bibinfo {volume} {98}},\ \bibinfo
  {pages} {235148} (\bibinfo {year} {2018})}\BibitemShut {NoStop}%
\bibitem [{\citenamefont {Xie}\ \emph {et~al.}(2017)\citenamefont {Xie},
  \citenamefont {Liao}, \citenamefont {Huang}, \citenamefont {Xie},
  \citenamefont {Chen}, \citenamefont {Liu},\ and\ \citenamefont
  {Xiang}}]{xie17}%
  \BibitemOpen
  \bibfield  {author} {\bibinfo {author} {\bibfnamefont {Z.~Y.}\ \bibnamefont
  {Xie}}, \bibinfo {author} {\bibfnamefont {H.~J.}\ \bibnamefont {Liao}},
  \bibinfo {author} {\bibfnamefont {R.~Z.}\ \bibnamefont {Huang}}, \bibinfo
  {author} {\bibfnamefont {H.~D.}\ \bibnamefont {Xie}}, \bibinfo {author}
  {\bibfnamefont {J.}~\bibnamefont {Chen}}, \bibinfo {author} {\bibfnamefont
  {Z.~Y.}\ \bibnamefont {Liu}},\ and\ \bibinfo {author} {\bibfnamefont
  {T.}~\bibnamefont {Xiang}},\ }\bibfield  {title} {\bibinfo {title} {Optimized
  contraction scheme for tensor-network states},\ }\href
  {https://doi.org/10.1103/PhysRevB.96.045128} {\bibfield  {journal} {\bibinfo
  {journal} {Phys. Rev. B}\ }\textbf {\bibinfo {volume} {96}},\ \bibinfo
  {pages} {045128} (\bibinfo {year} {2017})}\BibitemShut {NoStop}%
\bibitem [{\citenamefont {Haghshenas}\ \emph {et~al.}(2019)\citenamefont
  {Haghshenas}, \citenamefont {Gong},\ and\ \citenamefont
  {Sheng}}]{haghshenas19}%
  \BibitemOpen
  \bibfield  {author} {\bibinfo {author} {\bibfnamefont {R.}~\bibnamefont
  {Haghshenas}}, \bibinfo {author} {\bibfnamefont {S.-S.}\ \bibnamefont
  {Gong}},\ and\ \bibinfo {author} {\bibfnamefont {D.~N.}\ \bibnamefont
  {Sheng}},\ }\bibfield  {title} {\bibinfo {title} {Single-layer tensor network
  study of the {Heisenberg} model with chiral interactions on a kagome
  lattice},\ }\href {https://doi.org/10.1103/PhysRevB.99.174423} {\bibfield
  {journal} {\bibinfo  {journal} {Phys. Rev. B}\ }\textbf {\bibinfo {volume}
  {99}},\ \bibinfo {pages} {174423} (\bibinfo {year} {2019})}\BibitemShut
  {NoStop}%
\bibitem [{\citenamefont {Lan}\ and\ \citenamefont {Evenbly}(2023)}]{lan23}%
  \BibitemOpen
  \bibfield  {author} {\bibinfo {author} {\bibfnamefont {W.}~\bibnamefont
  {Lan}}\ and\ \bibinfo {author} {\bibfnamefont {G.}~\bibnamefont {Evenbly}},\
  }\bibfield  {title} {\bibinfo {title} {Reduced {Contraction} {Costs} of
  {Corner}-{Transfer} {Methods} for {PEPS}},\ }\href
  {http://arxiv.org/abs/2306.08212} {\bibfield  {journal} {\bibinfo  {journal}
  {arXiv:2306.08212 [quant-ph]}\ } (\bibinfo {year} {2023})}\BibitemShut
  {NoStop}%
\bibitem [{\citenamefont {Naumann}\ \emph {et~al.}(2025)\citenamefont
  {Naumann}, \citenamefont {Weerda}, \citenamefont {Eisert}, \citenamefont
  {Rizzi},\ and\ \citenamefont {Schmoll}}]{naumann25}%
  \BibitemOpen
  \bibfield  {author} {\bibinfo {author} {\bibfnamefont {J.}~\bibnamefont
  {Naumann}}, \bibinfo {author} {\bibfnamefont {E.~L.}\ \bibnamefont {Weerda}},
  \bibinfo {author} {\bibfnamefont {J.}~\bibnamefont {Eisert}}, \bibinfo
  {author} {\bibfnamefont {M.}~\bibnamefont {Rizzi}},\ and\ \bibinfo {author}
  {\bibfnamefont {P.}~\bibnamefont {Schmoll}},\ }\bibfield  {title} {\bibinfo
  {title} {Variationally optimizing infinite projected entangled-pair states at
  large bond dimensions: {A} split corner transfer matrix renormalization group
  approach},\ }\href {https://doi.org/10.1103/PhysRevB.111.235116} {\bibfield
  {journal} {\bibinfo  {journal} {Phys. Rev. B}\ }\textbf {\bibinfo {volume}
  {111}},\ \bibinfo {pages} {235116} (\bibinfo {year} {2025})}\BibitemShut
  {NoStop}%
\bibitem [{\citenamefont {Richards}\ and\ \citenamefont
  {S{\o}rensen}(2025)}]{richards25}%
  \BibitemOpen
  \bibfield  {author} {\bibinfo {author} {\bibfnamefont {A.~D.~S.}\
  \bibnamefont {Richards}}\ and\ \bibinfo {author} {\bibfnamefont {E.~S.}\
  \bibnamefont {S{\o}rensen}},\ }\bibfield  {title} {\bibinfo {title}
  {Ace-{TN}: {GPU}-{Accelerated} {Corner}-{Transfer}-{Matrix} {Renormalization}
  of {Infinite} {Projected} {Entangled}-{Pair} {States}},\ }\bibfield
  {journal} {\bibinfo  {journal} {arXiv:2503.13900 [cond-mat]}\ }\href
  {https://doi.org/10.48550/arXiv.2503.13900} {10.48550/arXiv.2503.13900}
  (\bibinfo {year} {2025})\BibitemShut {NoStop}%
\bibitem [{\citenamefont {Unfried}\ \emph {et~al.}(2023)\citenamefont
  {Unfried}, \citenamefont {Hauschild},\ and\ \citenamefont
  {Pollmann}}]{unfried23}%
  \BibitemOpen
  \bibfield  {author} {\bibinfo {author} {\bibfnamefont {J.}~\bibnamefont
  {Unfried}}, \bibinfo {author} {\bibfnamefont {J.}~\bibnamefont {Hauschild}},\
  and\ \bibinfo {author} {\bibfnamefont {F.}~\bibnamefont {Pollmann}},\
  }\bibfield  {title} {\bibinfo {title} {Fast time evolution of matrix product
  states using the {QR} decomposition},\ }\href
  {https://doi.org/10.1103/PhysRevB.107.155133} {\bibfield  {journal} {\bibinfo
   {journal} {Phys. Rev. B}\ }\textbf {\bibinfo {volume} {107}},\ \bibinfo
  {pages} {155133} (\bibinfo {year} {2023})}\BibitemShut {NoStop}%
\bibitem [{Note1()}]{Note1}%
  \BibitemOpen
  \bibinfo {note} {In the $C_{4v}$ symmetric case, the number of variational
  parameters is $n(D) = d/8 (D^4 + 2D^3 + 3D^2 + 2D)$}\BibitemShut {NoStop}%
\bibitem [{Note2()}]{Note2}%
  \BibitemOpen
  \bibinfo {note} {Here we initialize the boundary from the bulk tensors by
  contracting the corresponding bra- and ket- legs on the open
  boundary}\BibitemShut {NoStop}%
\bibitem [{Note3()}]{Note3}%
  \BibitemOpen
  \bibinfo {note} {Another type of reduced-rank approximation was introduced in
  Ref.~\cite {lan23} to decrease the computational cost. A reduced environment
  was also used in the directional CTMRG algorithm in Ref.~\cite {orus12} for
  the time evolution of a 1D quantum state, and in combination with a uniform
  MPS gauging method within a fixed-point CTMRG method~\cite
  {fishman18}}\BibitemShut {NoStop}%
\bibitem [{\citenamefont {Yang}(2025)}]{GithubQRCTM}%
  \BibitemOpen
  \bibfield  {author} {\bibinfo {author} {\bibfnamefont {Q.}~\bibnamefont
  {Yang}},\ }\href@noop {} {\bibinfo {title} {{QRCTM}}},\ \bibinfo
  {howpublished} {\url{https://github.com/qiyang-ustc/QRCTM}} (\bibinfo {year}
  {2025}),\ \bibinfo {note} {{GitHub} repository, main branch}\BibitemShut
  {NoStop}%
\bibitem [{\citenamefont {Jiang}\ \emph {et~al.}(2008)\citenamefont {Jiang},
  \citenamefont {Weng},\ and\ \citenamefont {Xiang}}]{jiang2008}%
  \BibitemOpen
  \bibfield  {author} {\bibinfo {author} {\bibfnamefont {H.~C.}\ \bibnamefont
  {Jiang}}, \bibinfo {author} {\bibfnamefont {Z.~Y.}\ \bibnamefont {Weng}},\
  and\ \bibinfo {author} {\bibfnamefont {T.}~\bibnamefont {Xiang}},\ }\bibfield
   {title} {\bibinfo {title} {{Accurate Determination of Tensor Network State
  of Quantum Lattice Models in Two Dimensions}},\ }\href
  {https://doi.org/10.1103/PhysRevLett.101.090603} {\bibfield  {journal}
  {\bibinfo  {journal} {Phys. Rev. Lett.}\ }\textbf {\bibinfo {volume} {101}},\
  \bibinfo {pages} {090603} (\bibinfo {year} {2008})}\BibitemShut {NoStop}%
\bibitem [{\citenamefont {Phien}\ \emph {et~al.}(2015)\citenamefont {Phien},
  \citenamefont {Bengua}, \citenamefont {Tuan}, \citenamefont {Corboz},\ and\
  \citenamefont {Orus}}]{phien15}%
  \BibitemOpen
  \bibfield  {author} {\bibinfo {author} {\bibfnamefont {H.~N.}\ \bibnamefont
  {Phien}}, \bibinfo {author} {\bibfnamefont {J.~A.}\ \bibnamefont {Bengua}},
  \bibinfo {author} {\bibfnamefont {H.~D.}\ \bibnamefont {Tuan}}, \bibinfo
  {author} {\bibfnamefont {P.}~\bibnamefont {Corboz}},\ and\ \bibinfo {author}
  {\bibfnamefont {R.}~\bibnamefont {Orus}},\ }\bibfield  {title} {\bibinfo
  {title} {Infinite projected entangled pair states algorithm improved: {Fast}
  full update and gauge fixing},\ }\href
  {https://doi.org/10.1103/PhysRevB.92.035142} {\bibfield  {journal} {\bibinfo
  {journal} {Phys. Rev. B}\ }\textbf {\bibinfo {volume} {92}},\ \bibinfo
  {pages} {035142} (\bibinfo {year} {2015})}\BibitemShut {NoStop}%
\bibitem [{\citenamefont {Corboz}(2016)}]{corboz16b}%
  \BibitemOpen
  \bibfield  {author} {\bibinfo {author} {\bibfnamefont {P.}~\bibnamefont
  {Corboz}},\ }\bibfield  {title} {\bibinfo {title} {Variational optimization
  with infinite projected entangled-pair states},\ }\href
  {https://doi.org/10.1103/PhysRevB.94.035133} {\bibfield  {journal} {\bibinfo
  {journal} {Phys. Rev. B}\ }\textbf {\bibinfo {volume} {94}},\ \bibinfo
  {pages} {035133} (\bibinfo {year} {2016})}\BibitemShut {NoStop}%
\bibitem [{\citenamefont {Vanderstraeten}\ \emph {et~al.}(2016)\citenamefont
  {Vanderstraeten}, \citenamefont {Haegeman}, \citenamefont {Corboz},\ and\
  \citenamefont {Verstraete}}]{vanderstraeten16}%
  \BibitemOpen
  \bibfield  {author} {\bibinfo {author} {\bibfnamefont {L.}~\bibnamefont
  {Vanderstraeten}}, \bibinfo {author} {\bibfnamefont {J.}~\bibnamefont
  {Haegeman}}, \bibinfo {author} {\bibfnamefont {P.}~\bibnamefont {Corboz}},\
  and\ \bibinfo {author} {\bibfnamefont {F.}~\bibnamefont {Verstraete}},\
  }\bibfield  {title} {\bibinfo {title} {Gradient methods for variational
  optimization of projected entangled-pair states},\ }\href
  {https://doi.org/10.1103/PhysRevB.94.155123} {\bibfield  {journal} {\bibinfo
  {journal} {Phys. Rev. B}\ }\textbf {\bibinfo {volume} {94}},\ \bibinfo
  {pages} {155123} (\bibinfo {year} {2016})}\BibitemShut {NoStop}%
\bibitem [{\citenamefont {Liao}\ \emph {et~al.}(2019)\citenamefont {Liao},
  \citenamefont {Liu}, \citenamefont {Wang},\ and\ \citenamefont
  {Xiang}}]{liao19}%
  \BibitemOpen
  \bibfield  {author} {\bibinfo {author} {\bibfnamefont {H.-J.}\ \bibnamefont
  {Liao}}, \bibinfo {author} {\bibfnamefont {J.-G.}\ \bibnamefont {Liu}},
  \bibinfo {author} {\bibfnamefont {L.}~\bibnamefont {Wang}},\ and\ \bibinfo
  {author} {\bibfnamefont {T.}~\bibnamefont {Xiang}},\ }\bibfield  {title}
  {\bibinfo {title} {Differentiable {Programming} {Tensor} {Networks}},\ }\href
  {https://doi.org/10.1103/PhysRevX.9.031041} {\bibfield  {journal} {\bibinfo
  {journal} {Phys. Rev. X}\ }\textbf {\bibinfo {volume} {9}},\ \bibinfo {pages}
  {031041} (\bibinfo {year} {2019})}\BibitemShut {NoStop}%
\bibitem [{\citenamefont {Francuz}\ \emph {et~al.}(2025)\citenamefont
  {Francuz}, \citenamefont {Schuch},\ and\ \citenamefont
  {Vanhecke}}]{francuz25}%
  \BibitemOpen
  \bibfield  {author} {\bibinfo {author} {\bibfnamefont {A.}~\bibnamefont
  {Francuz}}, \bibinfo {author} {\bibfnamefont {N.}~\bibnamefont {Schuch}},\
  and\ \bibinfo {author} {\bibfnamefont {B.}~\bibnamefont {Vanhecke}},\
  }\bibfield  {title} {\bibinfo {title} {Stable and efficient differentiation
  of tensor network algorithms},\ }\href
  {https://doi.org/10.1103/PhysRevResearch.7.013237} {\bibfield  {journal}
  {\bibinfo  {journal} {Phys. Rev. Res.}\ }\textbf {\bibinfo {volume} {7}},\
  \bibinfo {pages} {013237} (\bibinfo {year} {2025})}\BibitemShut {NoStop}%
\bibitem [{\citenamefont {Sandvik}\ and\ \citenamefont
  {Evertz}(2010)}]{Sandvik10}%
  \BibitemOpen
  \bibfield  {author} {\bibinfo {author} {\bibfnamefont {A.~W.}\ \bibnamefont
  {Sandvik}}\ and\ \bibinfo {author} {\bibfnamefont {H.~G.}\ \bibnamefont
  {Evertz}},\ }\bibfield  {title} {\bibinfo {title} {Loop updates for
  variational and projector quantum {Monte} {Carlo} simulations in the
  valence-bond basis},\ }\href {https://doi.org/10.1103/PhysRevB.82.024407}
  {\bibfield  {journal} {\bibinfo  {journal} {Phys. Rev. B}\ }\textbf {\bibinfo
  {volume} {82}},\ \bibinfo {pages} {024407} (\bibinfo {year}
  {2010})}\BibitemShut {NoStop}%
\bibitem [{\citenamefont {Hasik}(2024)}]{hasik24b}%
  \BibitemOpen
  \bibfield  {author} {\bibinfo {author} {\bibfnamefont {J.}~\bibnamefont
  {Hasik}},\ }\href {https://github.com/jurajHasik/j1j2_ipeps_states} {\bibinfo
  {title} {{jurajHasik}/j1j2\_ipeps\_states}} (\bibinfo {year} {2024}),\
  \bibinfo {note}
  {https://github.com/jurajHasik/j1j2\_ipeps\_states}\BibitemShut {NoStop}%
\bibitem [{\citenamefont {Hu}\ \emph {et~al.}(2013)\citenamefont {Hu},
  \citenamefont {Becca}, \citenamefont {Parola},\ and\ \citenamefont
  {Sorella}}]{hu13}%
  \BibitemOpen
  \bibfield  {author} {\bibinfo {author} {\bibfnamefont {W.-J.}\ \bibnamefont
  {Hu}}, \bibinfo {author} {\bibfnamefont {F.}~\bibnamefont {Becca}}, \bibinfo
  {author} {\bibfnamefont {A.}~\bibnamefont {Parola}},\ and\ \bibinfo {author}
  {\bibfnamefont {S.}~\bibnamefont {Sorella}},\ }\bibfield  {title} {\bibinfo
  {title} {Direct evidence for a gapless ${{Z}}_{2}$ spin liquid by frustrating
  {N}\'eel antiferromagnetism},\ }\href
  {https://doi.org/10.1103/PhysRevB.88.060402} {\bibfield  {journal} {\bibinfo
  {journal} {Phys. Rev. B}\ }\textbf {\bibinfo {volume} {88}},\ \bibinfo
  {pages} {060402} (\bibinfo {year} {2013})}\BibitemShut {NoStop}%
\bibitem [{\citenamefont {Richter}\ \emph {et~al.}(2015)\citenamefont
  {Richter}, \citenamefont {Zinke},\ and\ \citenamefont {Farnell}}]{richter15}%
  \BibitemOpen
  \bibfield  {author} {\bibinfo {author} {\bibfnamefont {J.}~\bibnamefont
  {Richter}}, \bibinfo {author} {\bibfnamefont {R.}~\bibnamefont {Zinke}},\
  and\ \bibinfo {author} {\bibfnamefont {D.~J.~J.}\ \bibnamefont {Farnell}},\
  }\bibfield  {title} {\bibinfo {title} {The spin-1/2 square-lattice {J1}-{J2}
  model: the spin-gap issue},\ }\href
  {https://doi.org/10.1140/epjb/e2014-50589-x} {\bibfield  {journal} {\bibinfo
  {journal} {Eur. Phys. J. B}\ }\textbf {\bibinfo {volume} {88}},\ \bibinfo
  {pages} {2} (\bibinfo {year} {2015})}\BibitemShut {NoStop}%
\bibitem [{\citenamefont {Wang}\ and\ \citenamefont {Sandvik}(2018)}]{wang18}%
  \BibitemOpen
  \bibfield  {author} {\bibinfo {author} {\bibfnamefont {L.}~\bibnamefont
  {Wang}}\ and\ \bibinfo {author} {\bibfnamefont {A.~W.}\ \bibnamefont
  {Sandvik}},\ }\bibfield  {title} {\bibinfo {title} {Critical {Level}
  {Crossings} and {Gapless} {Spin} {Liquid} in the {Square}-{Lattice}
  {Spin}-$1/2$ ${{J}}_{1}{-}{{J}}_{2}$ {Heisenberg} {Antiferromagnet}},\ }\href
  {https://doi.org/10.1103/PhysRevLett.121.107202} {\bibfield  {journal}
  {\bibinfo  {journal} {Phys. Rev. Lett.}\ }\textbf {\bibinfo {volume} {121}},\
  \bibinfo {pages} {107202} (\bibinfo {year} {2018})}\BibitemShut {NoStop}%
\bibitem [{\citenamefont {Hering}\ \emph {et~al.}(2019)\citenamefont {Hering},
  \citenamefont {Sonnenschein}, \citenamefont {Iqbal},\ and\ \citenamefont
  {Reuther}}]{hering19}%
  \BibitemOpen
  \bibfield  {author} {\bibinfo {author} {\bibfnamefont {M.}~\bibnamefont
  {Hering}}, \bibinfo {author} {\bibfnamefont {J.}~\bibnamefont
  {Sonnenschein}}, \bibinfo {author} {\bibfnamefont {Y.}~\bibnamefont
  {Iqbal}},\ and\ \bibinfo {author} {\bibfnamefont {J.}~\bibnamefont
  {Reuther}},\ }\bibfield  {title} {\bibinfo {title} {Characterization of
  quantum spin liquids and their spinon band structures via functional
  renormalization},\ }\href {https://doi.org/10.1103/PhysRevB.99.100405}
  {\bibfield  {journal} {\bibinfo  {journal} {Phys. Rev. B}\ }\textbf {\bibinfo
  {volume} {99}},\ \bibinfo {pages} {100405} (\bibinfo {year}
  {2019})}\BibitemShut {NoStop}%
\bibitem [{\citenamefont {Nomura}\ and\ \citenamefont
  {Imada}(2021)}]{nomura21}%
  \BibitemOpen
  \bibfield  {author} {\bibinfo {author} {\bibfnamefont {Y.}~\bibnamefont
  {Nomura}}\ and\ \bibinfo {author} {\bibfnamefont {M.}~\bibnamefont {Imada}},\
  }\bibfield  {title} {\bibinfo {title} {Dirac-{Type} {Nodal} {Spin} {Liquid}
  {Revealed} by {Refined} {Quantum} {Many}-{Body} {Solver} {Using}
  {Neural}-{Network} {Wave} {Function}, {Correlation} {Ratio}, and {Level}
  {Spectroscopy}},\ }\href {https://doi.org/10.1103/PhysRevX.11.031034}
  {\bibfield  {journal} {\bibinfo  {journal} {Phys. Rev. X}\ }\textbf {\bibinfo
  {volume} {11}},\ \bibinfo {pages} {031034} (\bibinfo {year}
  {2021})}\BibitemShut {NoStop}%
\bibitem [{\citenamefont {Liu}\ \emph {et~al.}(2022{\natexlab{b}})\citenamefont
  {Liu}, \citenamefont {Gong}, \citenamefont {Li}, \citenamefont {Poilblanc},
  \citenamefont {Chen},\ and\ \citenamefont {Gu}}]{liu22}%
  \BibitemOpen
  \bibfield  {author} {\bibinfo {author} {\bibfnamefont {W.-Y.}\ \bibnamefont
  {Liu}}, \bibinfo {author} {\bibfnamefont {S.-S.}\ \bibnamefont {Gong}},
  \bibinfo {author} {\bibfnamefont {Y.-B.}\ \bibnamefont {Li}}, \bibinfo
  {author} {\bibfnamefont {D.}~\bibnamefont {Poilblanc}}, \bibinfo {author}
  {\bibfnamefont {W.-Q.}\ \bibnamefont {Chen}},\ and\ \bibinfo {author}
  {\bibfnamefont {Z.-C.}\ \bibnamefont {Gu}},\ }\bibfield  {title} {\bibinfo
  {title} {Gapless quantum spin liquid and global phase diagram of the spin-1/2
  {J1}-{J2} square antiferromagnetic {Heisenberg} model},\ }\href
  {https://doi.org/10.1016/j.scib.2022.03.010} {\bibfield  {journal} {\bibinfo
  {journal} {Science Bulletin}\ }\textbf {\bibinfo {volume} {67}},\ \bibinfo
  {pages} {1034} (\bibinfo {year} {2022}{\natexlab{b}})}\BibitemShut {NoStop}%
\bibitem [{\citenamefont {Dong}\ \emph {et~al.}(2025)\citenamefont {Dong},
  \citenamefont {Wang}, \citenamefont {Zhang}, \citenamefont {Zhang},\ and\
  \citenamefont {He}}]{dong25}%
  \BibitemOpen
  \bibfield  {author} {\bibinfo {author} {\bibfnamefont {S.}~\bibnamefont
  {Dong}}, \bibinfo {author} {\bibfnamefont {C.}~\bibnamefont {Wang}}, \bibinfo
  {author} {\bibfnamefont {H.}~\bibnamefont {Zhang}}, \bibinfo {author}
  {\bibfnamefont {M.}~\bibnamefont {Zhang}},\ and\ \bibinfo {author}
  {\bibfnamefont {L.}~\bibnamefont {He}},\ }\bibfield  {title} {\bibinfo
  {title} {Efficient {Projected} {Entangled} {Pair} {States} {Methods} for
  {Periodic} {Quantum} {Systems}},\ }\href {https://doi.org/10.1103/fvq2-msr4}
  {\bibfield  {journal} {\bibinfo  {journal} {Phys. Rev. Lett.}\ }\textbf
  {\bibinfo {volume} {135}},\ \bibinfo {pages} {026501} (\bibinfo {year}
  {2025})}\BibitemShut {NoStop}%
\bibitem [{\citenamefont {Liu}\ \emph {et~al.}(2024)\citenamefont {Liu},
  \citenamefont {Du}, \citenamefont {Peng}, \citenamefont {Gray},\ and\
  \citenamefont {Chan}}]{liu24b}%
  \BibitemOpen
  \bibfield  {author} {\bibinfo {author} {\bibfnamefont {W.-Y.}\ \bibnamefont
  {Liu}}, \bibinfo {author} {\bibfnamefont {S.-J.}\ \bibnamefont {Du}},
  \bibinfo {author} {\bibfnamefont {R.}~\bibnamefont {Peng}}, \bibinfo {author}
  {\bibfnamefont {J.}~\bibnamefont {Gray}},\ and\ \bibinfo {author}
  {\bibfnamefont {G.~K.-L.}\ \bibnamefont {Chan}},\ }\bibfield  {title}
  {\bibinfo {title} {Tensor {Network} {Computations} {That} {Capture} {Strict}
  {Variationality}, {Volume} {Law} {Behavior}, and the {Efficient}
  {Representation} of {Neural} {Network} {States}},\ }\href
  {https://doi.org/10.1103/PhysRevLett.133.260404} {\bibfield  {journal}
  {\bibinfo  {journal} {Phys. Rev. Lett.}\ }\textbf {\bibinfo {volume} {133}},\
  \bibinfo {pages} {260404} (\bibinfo {year} {2024})}\BibitemShut {NoStop}%
\bibitem [{\citenamefont {Zhao}\ \emph {et~al.}(2022)\citenamefont {Zhao},
  \citenamefont {Li}, \citenamefont {Xiao}, \citenamefont {Chen}, \citenamefont
  {Wang}, \citenamefont {Shen}, \citenamefont {Zhao}, \citenamefont {Wu},
  \citenamefont {An}, \citenamefont {He},\ and\ \citenamefont
  {Liang}}]{zhao22}%
  \BibitemOpen
  \bibfield  {author} {\bibinfo {author} {\bibfnamefont {X.}~\bibnamefont
  {Zhao}}, \bibinfo {author} {\bibfnamefont {M.}~\bibnamefont {Li}}, \bibinfo
  {author} {\bibfnamefont {Q.}~\bibnamefont {Xiao}}, \bibinfo {author}
  {\bibfnamefont {J.}~\bibnamefont {Chen}}, \bibinfo {author} {\bibfnamefont
  {F.}~\bibnamefont {Wang}}, \bibinfo {author} {\bibfnamefont {L.}~\bibnamefont
  {Shen}}, \bibinfo {author} {\bibfnamefont {M.}~\bibnamefont {Zhao}}, \bibinfo
  {author} {\bibfnamefont {W.}~\bibnamefont {Wu}}, \bibinfo {author}
  {\bibfnamefont {H.}~\bibnamefont {An}}, \bibinfo {author} {\bibfnamefont
  {L.}~\bibnamefont {He}},\ and\ \bibinfo {author} {\bibfnamefont
  {X.}~\bibnamefont {Liang}},\ }\bibfield  {title} {\bibinfo {title} {{ AI for
  Quantum Mechanics: High Performance Quantum Many-Body Simulations via Deep
  Learning }},\ }in\ \href {https://doi.org/10.1109/SC41404.2022.00053} {\emph
  {\bibinfo {booktitle} {SC22: International Conference for High Performance
  Computing, Networking, Storage and Analysis}}}\ (\bibinfo  {publisher} {IEEE
  Computer Society},\ \bibinfo {address} {Los Alamitos, CA, USA},\ \bibinfo
  {year} {2022})\ pp.\ \bibinfo {pages} {1--15}\BibitemShut {NoStop}%
\bibitem [{\citenamefont {Liang}\ \emph {et~al.}(2023)\citenamefont {Liang},
  \citenamefont {Li}, \citenamefont {Xiao}, \citenamefont {Chen}, \citenamefont
  {Yang}, \citenamefont {An},\ and\ \citenamefont {He}}]{liang23}%
  \BibitemOpen
  \bibfield  {author} {\bibinfo {author} {\bibfnamefont {X.}~\bibnamefont
  {Liang}}, \bibinfo {author} {\bibfnamefont {M.}~\bibnamefont {Li}}, \bibinfo
  {author} {\bibfnamefont {Q.}~\bibnamefont {Xiao}}, \bibinfo {author}
  {\bibfnamefont {J.}~\bibnamefont {Chen}}, \bibinfo {author} {\bibfnamefont
  {C.}~\bibnamefont {Yang}}, \bibinfo {author} {\bibfnamefont {H.}~\bibnamefont
  {An}},\ and\ \bibinfo {author} {\bibfnamefont {L.}~\bibnamefont {He}},\
  }\bibfield  {title} {\bibinfo {title} {Deep learning representations for
  quantum many-body systems on heterogeneous hardware},\ }\href
  {https://doi.org/10.1088/2632-2153/acc56a} {\bibfield  {journal} {\bibinfo
  {journal} {Mach. Learn.: Sci. Technol.}\ }\textbf {\bibinfo {volume} {4}},\
  \bibinfo {pages} {015035} (\bibinfo {year} {2023})}\BibitemShut {NoStop}%
\bibitem [{\citenamefont {Roth}\ \emph {et~al.}(2023)\citenamefont {Roth},
  \citenamefont {Szab{\'o}},\ and\ \citenamefont {MacDonald}}]{roth23}%
  \BibitemOpen
  \bibfield  {author} {\bibinfo {author} {\bibfnamefont {C.}~\bibnamefont
  {Roth}}, \bibinfo {author} {\bibfnamefont {A.}~\bibnamefont {Szab{\'o}}},\
  and\ \bibinfo {author} {\bibfnamefont {A.~H.}\ \bibnamefont {MacDonald}},\
  }\bibfield  {title} {\bibinfo {title} {High-accuracy variational {Monte}
  {Carlo} for frustrated magnets with deep neural networks},\ }\href
  {https://doi.org/10.1103/PhysRevB.108.054410} {\bibfield  {journal} {\bibinfo
   {journal} {Phys. Rev. B}\ }\textbf {\bibinfo {volume} {108}},\ \bibinfo
  {pages} {054410} (\bibinfo {year} {2023})}\BibitemShut {NoStop}%
\bibitem [{\citenamefont {Chen}\ and\ \citenamefont {Heyl}(2024)}]{chen24}%
  \BibitemOpen
  \bibfield  {author} {\bibinfo {author} {\bibfnamefont {A.}~\bibnamefont
  {Chen}}\ and\ \bibinfo {author} {\bibfnamefont {M.}~\bibnamefont {Heyl}},\
  }\bibfield  {title} {\bibinfo {title} {Empowering deep neural quantum states
  through efficient optimization},\ }\href
  {https://doi.org/10.1038/s41567-024-02566-1} {\bibfield  {journal} {\bibinfo
  {journal} {Nat. Phys.}\ }\textbf {\bibinfo {volume} {20}},\ \bibinfo {pages}
  {1476} (\bibinfo {year} {2024})}\BibitemShut {NoStop}%
\bibitem [{\citenamefont {Rende}\ \emph {et~al.}(2024)\citenamefont {Rende},
  \citenamefont {Viteritti}, \citenamefont {Bardone}, \citenamefont {Becca},\
  and\ \citenamefont {Goldt}}]{rende24}%
  \BibitemOpen
  \bibfield  {author} {\bibinfo {author} {\bibfnamefont {R.}~\bibnamefont
  {Rende}}, \bibinfo {author} {\bibfnamefont {L.~L.}\ \bibnamefont
  {Viteritti}}, \bibinfo {author} {\bibfnamefont {L.}~\bibnamefont {Bardone}},
  \bibinfo {author} {\bibfnamefont {F.}~\bibnamefont {Becca}},\ and\ \bibinfo
  {author} {\bibfnamefont {S.}~\bibnamefont {Goldt}},\ }\bibfield  {title}
  {\bibinfo {title} {A simple linear algebra identity to optimize large-scale
  neural network quantum states},\ }\href
  {https://doi.org/10.1038/s42005-024-01732-4} {\bibfield  {journal} {\bibinfo
  {journal} {Commun Phys}\ }\textbf {\bibinfo {volume} {7}},\ \bibinfo {pages}
  {1} (\bibinfo {year} {2024})}\BibitemShut {NoStop}%
\bibitem [{Note4()}]{Note4}%
  \BibitemOpen
  \bibinfo {note} {The boundary bond dimensions used here are $\chi
  _{opt}=\{200, 200, 300, 300, 300\}$, $\chi _{eval}=\{200, 200, 600, 600,
  700\}$ for $D=\{4, 5, 6, 7, 8\}$, respectively.}\BibitemShut {Stop}%
\bibitem [{\citenamefont {Singh}\ \emph {et~al.}(2011)\citenamefont {Singh},
  \citenamefont {Pfeifer},\ and\ \citenamefont {Vidal}}]{singh2011}%
  \BibitemOpen
  \bibfield  {author} {\bibinfo {author} {\bibfnamefont {S.}~\bibnamefont
  {Singh}}, \bibinfo {author} {\bibfnamefont {R.~N.~C.}\ \bibnamefont
  {Pfeifer}},\ and\ \bibinfo {author} {\bibfnamefont {G.}~\bibnamefont
  {Vidal}},\ }\bibfield  {title} {\bibinfo {title} {{Tensor network states and
  algorithms in the presence of a global U(1) symmetry}},\ }\href
  {https://doi.org/10.1103/PhysRevB.83.115125} {\bibfield  {journal} {\bibinfo
  {journal} {Phys. Rev. B}\ }\textbf {\bibinfo {volume} {83}},\ \bibinfo
  {pages} {115125} (\bibinfo {year} {2011})}\BibitemShut {NoStop}%
\bibitem [{\citenamefont {Bauer}\ \emph {et~al.}(2011)\citenamefont {Bauer},
  \citenamefont {Corboz}, \citenamefont {Or\'us},\ and\ \citenamefont
  {Troyer}}]{bauer2011}%
  \BibitemOpen
  \bibfield  {author} {\bibinfo {author} {\bibfnamefont {B.}~\bibnamefont
  {Bauer}}, \bibinfo {author} {\bibfnamefont {P.}~\bibnamefont {Corboz}},
  \bibinfo {author} {\bibfnamefont {R.}~\bibnamefont {Or\'us}},\ and\ \bibinfo
  {author} {\bibfnamefont {M.}~\bibnamefont {Troyer}},\ }\bibfield  {title}
  {\bibinfo {title} {Implementing global abelian symmetries in projected
  entangled-pair state algorithms},\ }\href
  {https://doi.org/10.1103/PhysRevB.83.125106} {\bibfield  {journal} {\bibinfo
  {journal} {Phys. Rev. B}\ }\textbf {\bibinfo {volume} {83}},\ \bibinfo
  {pages} {125106} (\bibinfo {year} {2011})}\BibitemShut {NoStop}%
\bibitem [{\citenamefont {van Alphen}\ \emph {et~al.}(2025)\citenamefont {van
  Alphen}, \citenamefont {Kleijweg}, \citenamefont {Hasik},\ and\ \citenamefont
  {Corboz}}]{alphen25}%
  \BibitemOpen
  \bibfield  {author} {\bibinfo {author} {\bibfnamefont {O.}~\bibnamefont {van
  Alphen}}, \bibinfo {author} {\bibfnamefont {S.~V.}\ \bibnamefont {Kleijweg}},
  \bibinfo {author} {\bibfnamefont {J.}~\bibnamefont {Hasik}},\ and\ \bibinfo
  {author} {\bibfnamefont {P.}~\bibnamefont {Corboz}},\ }\bibfield  {title}
  {\bibinfo {title} {Exploiting the {Hermitian} symmetry in tensor network
  algorithms},\ }\href {https://doi.org/10.1103/PhysRevB.111.045105} {\bibfield
   {journal} {\bibinfo  {journal} {Phys. Rev. B}\ }\textbf {\bibinfo {volume}
  {111}},\ \bibinfo {pages} {045105} (\bibinfo {year} {2025})}\BibitemShut
  {NoStop}%
\bibitem [{\citenamefont {Czarnik}\ and\ \citenamefont
  {Dziarmaga}(2015)}]{czarnik15b}%
  \BibitemOpen
  \bibfield  {author} {\bibinfo {author} {\bibfnamefont {P.}~\bibnamefont
  {Czarnik}}\ and\ \bibinfo {author} {\bibfnamefont {J.}~\bibnamefont
  {Dziarmaga}},\ }\bibfield  {title} {\bibinfo {title} {Variational approach to
  projected entangled pair states at finite temperature},\ }\href
  {https://doi.org/10.1103/PhysRevB.92.035152} {\bibfield  {journal} {\bibinfo
  {journal} {Phys. Rev. B}\ }\textbf {\bibinfo {volume} {92}},\ \bibinfo
  {pages} {035152} (\bibinfo {year} {2015})}\BibitemShut {NoStop}%
\bibitem [{\citenamefont {Czarnik}\ \emph {et~al.}(2016)\citenamefont
  {Czarnik}, \citenamefont {Rams},\ and\ \citenamefont
  {Dziarmaga}}]{czarnik16b}%
  \BibitemOpen
  \bibfield  {author} {\bibinfo {author} {\bibfnamefont {P.}~\bibnamefont
  {Czarnik}}, \bibinfo {author} {\bibfnamefont {M.~M.}\ \bibnamefont {Rams}},\
  and\ \bibinfo {author} {\bibfnamefont {J.}~\bibnamefont {Dziarmaga}},\
  }\bibfield  {title} {\bibinfo {title} {Variational tensor network
  renormalization in imaginary time: {Benchmark} results in the {Hubbard} model
  at finite temperature},\ }\href {https://doi.org/10.1103/PhysRevB.94.235142}
  {\bibfield  {journal} {\bibinfo  {journal} {Phys. Rev. B}\ }\textbf {\bibinfo
  {volume} {94}},\ \bibinfo {pages} {235142} (\bibinfo {year}
  {2016})}\BibitemShut {NoStop}%
\bibitem [{\citenamefont {Czarnik}\ \emph {et~al.}(2019)\citenamefont
  {Czarnik}, \citenamefont {Dziarmaga},\ and\ \citenamefont
  {Corboz}}]{czarnik19}%
  \BibitemOpen
  \bibfield  {author} {\bibinfo {author} {\bibfnamefont {P.}~\bibnamefont
  {Czarnik}}, \bibinfo {author} {\bibfnamefont {J.}~\bibnamefont {Dziarmaga}},\
  and\ \bibinfo {author} {\bibfnamefont {P.}~\bibnamefont {Corboz}},\
  }\bibfield  {title} {\bibinfo {title} {Time evolution of an infinite
  projected entangled pair state: {An} efficient algorithm},\ }\href
  {https://doi.org/10.1103/PhysRevB.99.035115} {\bibfield  {journal} {\bibinfo
  {journal} {Phys. Rev. B}\ }\textbf {\bibinfo {volume} {99}},\ \bibinfo
  {pages} {035115} (\bibinfo {year} {2019})}\BibitemShut {NoStop}%
\bibitem [{\citenamefont {Or{\'u}s}(2012)}]{orus12}%
  \BibitemOpen
  \bibfield  {author} {\bibinfo {author} {\bibfnamefont {R.}~\bibnamefont
  {Or{\'u}s}},\ }\bibfield  {title} {\bibinfo {title} {Exploring corner
  transfer matrices and corner tensors for the classical simulation of quantum
  lattice systems},\ }\href {https://doi.org/10.1103/PhysRevB.85.205117}
  {\bibfield  {journal} {\bibinfo  {journal} {Phys. Rev. B}\ }\textbf {\bibinfo
  {volume} {85}},\ \bibinfo {pages} {205117} (\bibinfo {year}
  {2012})}\BibitemShut {NoStop}%
\end{thebibliography}%

\end{document}